\documentstyle[aas2pp4,psbox]{article}

\slugcomment{ApJ, accepted}
\lefthead{Imanishi, Koyama, \& Tsuboi}
\righthead {$Chandra$ Observation of the $\rho$ Oph cloud}

\begin{document}

\title{$Chandra$ Observation of the $\rho$ Oph Cloud}

\vspace{0.3mm}
\author{Kensuke~Imanishi, Katsuji~Koyama\altaffilmark{1}}
\affil{Department of Physics, Graduate School of Science, Kyoto
University, Sakyo-ku, Kyoto, 606-8502, Japan;
kensuke@cr.scphys.kyoto-u.ac.jp, koyama@cr.scphys.kyoto-u.ac.jp}

\and

\author{Yohko~Tsuboi}
\affil{Department of Astronomy \& Astrophysics, 525 Davey Laboratory,
Pennsylvania State University, University Park PA 16802, USA; 
tsuboi@astro.psu.edu}

\altaffiltext{1}{CREST, Japan Science and Technology Corporation (JST),
4-1-8 Honmachi, Kawaguchi, Saitama, 332-0012, Japan}

\begin{abstract}

We observed a 17$'\times$17$'$ region of the $\rho$ Oph molecular cloud,
using the ACIS-I array onboard the {\it Chandra X-ray Observatory}
(CXO). A 100-ks exposure observation revealed $\sim$ 100 X-ray sources
above the detection limit of $\sim$ 10$^{28}$ ergs s$^{-1}$.  About 2/3
of the X-ray sources are identified with an optical and/or infrared
counterpart, including significant numbers of Class I, II and III
sources and a few brown dwarfs. The X-ray detection rate of Class I
sources is a surprisingly high 70 \%. Many X-ray flares, possibly due to
the magnetic activity, are found from all of the classes.  We make
systematic spectral studies of the quiescent and flare X-rays for all
the X-ray sources.  The X-ray temperatures and absorptions of Class I
sources are generally larger than those of Class II and III sources.
Also X-ray flares from Class I sources show slightly higher duty ratio,
temperature, and luminosity, than those of more evolved classes. We
further give brief results on brown dwarfs, sources in a younger phase
than Class I, unclassified, and unidentified sources.  Details for
several selected young stellar objects,
%
%
including the first detection of a neutral iron line (6.4~keV) from a
Class I source (YLW 16A), are separately addressed.

\end{abstract}

\section{INTRODUCTION}

Evolution of low-mass young stellar objects (YSOs) can be traced by the
infrared -- sub $mm$ radio spectral energy distributions (SEDs):
molecular cloud cores through protostars are represented by Class 0 and
Class I SEDs, while classical T Tauri stars (CTTSs) and weak-line T
Tauri stars (WTTSs) exhibit Class II and Class III SEDs (Shu, Adams, \&
Lizano 1987; Andr\'e \& Montmerle 1994).  With the {\it Einstein}
satellite, T Tauri stars (TTSs = CTTSs and WTTSs) have been found to
emit soft X-rays, with the X-ray characteristics of moderate plasma
temperature ($\sim$ 1 keV), strong variability and occasional rapid
flares. These are consistent with the scenario of enhanced solar-type
activity, attributable to magnetic dynamo processes (Feigelson \&
DeCampli 1981; Montmerle et al. 1983).

Protostars are often embedded deeply in star forming clouds, shrouded by
dense circumstellar gas and dust, hence generally invisible in the
optical, near infrared and even soft X-ray bands. X-rays from Class I
sources (the late phase of protostars) have been reported in deep {\it
ASCA} and {\it ROSAT} images of the $\rho$ Ophiuchi cloud (Koyama et al. 
1994; Casanova et al. 1995; Kamata et al. 1997; Grosso et al. 1997;
Carkner, Kozak, \& Feigelson 1998; Tsuboi 1999; Tsuboi et al. 2000;
Grosso et al. 2000; Grosso 2001). However possible contamination from
nearby TTSs (or others) would not be removed with the limited spatial
resolution of {\it ASCA}. Very low count rates from {\it ROSAT} hampered
quantitative X-ray study on these Class I sources.  In order to shed
light on the embedded YSOs, like Class I sources in the $\rho$ Ophiuchi
cloud, we have observed this cloud with the {\it Chandra X-ray
Observatory} (CXO, Weisskopf, O$'$dell, \& van Speybroeck 1997), which
has wide energy band sensitivity coupled with unprecedented spatial
resolution.

In this paper, we systematically study the X-ray features of the young
stellar objects in the $\rho$ Ophiuchi cloud based on the $Chandra$
results.  For this purpose, we first clarify our criteria of source
classifications based on the near- to far-IR data.  Class I sources are
selected from near- to mid-IR band or near- to far-IR band spectra
(Andr\'e \& Montmerle 1994; Casanova et al. 1995; Chen et al. 1995; Chen
et al. 1997; Motte, Andr\'e, \& Neri 1998; Luhman \& Rieke 1999; Grosso
et al. 2000). Since the classifications of these papers are not fully
consistent with each other, we regard a bona fide Class I source (Class
I, here and after) if it is referred to as a Class I source by all the
papers, and the others to be Class I candidates (Class I$_c$, hear and
after). Class II and III sources (here Class II and Class III) are
simply selected by the spectral index between near- and mid- IR bands
(Andr\'e \& Montmerle 1994; Casanova et al. 1995; Luhman \& Rieke 1999;
Grosso et al. 2000).  The age or evolution sequence of Class I$_c$
should be in between Class I and Class II.  We also use the terminology
``Class I+I$_c$'' by combining the Class I and Class I$_c$ for
simplicity.  Likewise, ``Class II+III'' means a combined group of Class
II and Class III.  A brown dwarf (here and after, BD) means a star with
mass falling well below the hydrogen-burning limit of 0.08 M$_{\odot}$,
and a brown dwarf candidate (here and after, BD$_c$) indicates a star
with mass close to 0.08 M$_{\odot}$ (Wilking, Greene, \& Meyer 1999;
Cushing, Tokunaga, \& Kobayashi 2000).

For comparison with the previous X-ray and infrared results, we assume
the distance to the $\rho$ Oph region to be 165 pc (Dame et al. 1987)
throughout this paper, although $Hipparcos$ data suggest a closer
distance of $\sim$120 pc (Knude \& Hog 1998).
 
\section{OBSERVATION AND DATA REDUCTION}

The observation was made on 2000 April 13--14 with the ACIS-I array
consisting of four abutted X-ray CCDs, which covered a field including
the southeast part of the $\rho$ Oph cloud in the 0.2--10 keV band.  
%
%
The telescope optical axis position on the ACIS-I array is R.A. =
16$^{\rm h}$27$^{\rm m}$18.1$^{\rm s}$, Dec =
$-$24$^{\circ}$34$'$21.9$''$ (epoch 2000). We use Level 1 processed
events provided by the pipeline processing at the Chandra X-ray Center.
The total available exposure time is $\approx$ $10^2$ ks.  To minimize
the effect of the degradation of charge transfer inefficiency (CTI), we
apply the improved data processing technique developed by Townsley et
al. (2000).  To reject background events, we apply a grade filter to
keep only {\it ASCA} grades\footnote{see
http://asc.harvard.edu/udocs/docs/POG/MPOG/index.html} 0, 2, 3, 4, and
6.

\section{ANALYSIS AND RESULTS}

\subsection{Source Detection}

Figure 1 shows the ACIS-I false-color image of the $\rho$ Oph cloud. Red
and blue colors represent photons in the soft (0.7--2.0 keV) and hard
(2.0--7.0 keV) X-ray bands, respectively. For source finding, we run the
program {\it wavdetect}\footnote{see
http://asc.harvard.edu/udocs/docs/swdocs/detect/html/} in the total
(0.5--9.0 keV) X-ray band. We set the significance criterion at
10$^{-7}$ and wavelet scales ranging from 1 to 16 pixels in multiples of
$\sqrt{2}$. In the 17$'$.4$\times$17$'$.4 field, we detected 87 sources
%
%
with a confidence level of $\gtrsim$5.3 $\sigma$.  The source position and
ACIS-I count rate for each source are listed in Table 1 (columns 2--5),
where the X-ray photons are extracted from 1$''$.5--22$''$.7 radius
circles, depending on the angular distance from the optical axis of the
telescope.

\begin{figure}
 \psbox[xsize=0.9\textwidth]{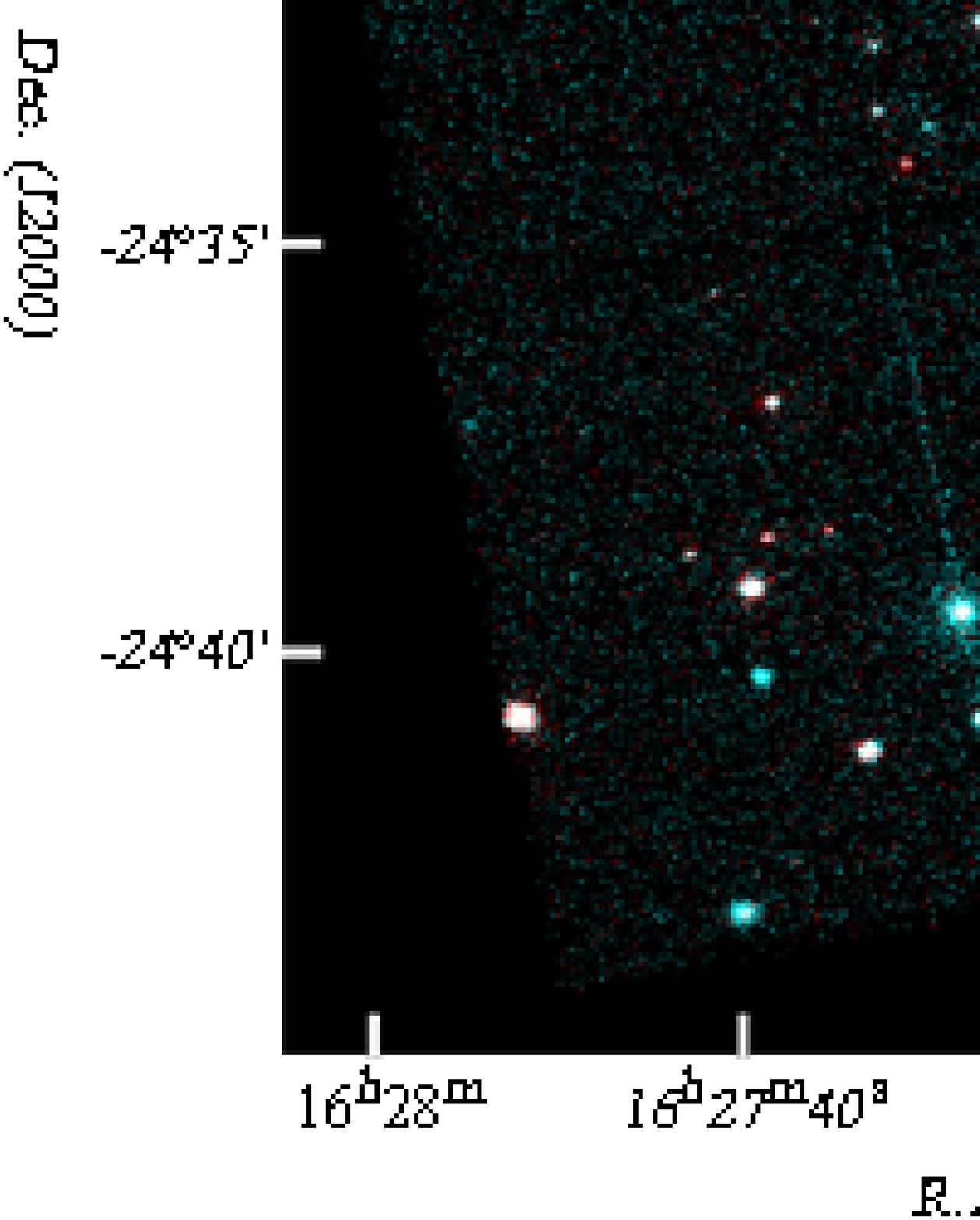}

 \caption[f1.eps]{ACIS-I image of the $\rho$ Oph cloud. Red and blue
colors represent photons in the soft (0.7--2.0 keV) and hard (2.0--7.0
keV) X-ray bands, respectively. \label{fig:image} }
\end{figure}

\placefigure{fig:image}

\placetable{tab:target}

\subsection{Source Classification}

We searched for a near-infrared counterpart from the three-color ($JHK$)
imaging survey by Barsony et al. (1997), and find that 2/3 (58) of the
X-ray sources have a counterpart (column 2 in Table 2). The other 1/3
(29) have no counterpart (here and after, unidentified sources), most of
which are blue ( hard X-ray sources) in Figure 1.  Using the criteria
given in \S1, we find 7 Class Is and 11 Class I$_c$s; most of them
appeared blue in Figure 1, or mainly emit hard X-ray
photons. We also find 1 BD and 1 BD$_c$ (No.76 and No.81), and at least
1 foreground star (No.52, Festin 1998).  The other 37 sources are mainly
colored red, hence are soft band sources, of which 20 are classified as
Class II+IIIs, and 17 are unclassified yet (here and after, unclassified
sources).

%
%
We also search for counterparts in the previous X-ray surveys with
$ROSAT$ PSPC (Casanova et al. 1995), $ROSAT$ HRI (Grosso et al. 2000;
Grosso 2001), and $ASCA$ (Kamata et al. 1997; Tsuboi et al. 2000), and
find only 22 sources in these surveys (columns 3--5 in Table 2).  In
contrast, while most PSPC and $ASCA$ sources are found in the table,
about 10 HRI sources in our field of view are not detected, although the
typical HRI luminosity of 10$^{30}$ ergs s$^{-1}$ is well above our
detection limit of $\sim$ 10$^{28}$ ergs s$^{-1}$ (see \S4).  The
results of the classifications and identifications are summarized in
Table 2.

\placetable{tab:id}

\subsection{Timing Analysis and  Flares}

For all the X-ray sources, we make light curves in the total X-ray band.
We define a flare by following two criteria: (1) a constant flux
hypothesis for the light curve is rejected with a chi-square fitting,
and (2) the X-ray flux increases and/or decreases by more than double
the preceding/following flux with a time scale smaller than a few 10$^4$
sec.

We then pick up 7, 7, 12, 9, and 1 flares from 7 class Is, 11 Class
I$_c$s, 20 Class II+IIIs, 17 unclassified sources, and 29 unidentified
sources, respectively.  For example, the flare light curves of Class
I+I$_c$s are given in Figure 2.  A Class I source YLW 16A (No.57)
exhibited a giant flare, in which a significant fraction of X-ray
photons pile up (more than two X-ray photons register in a single ACIS
pixel within one read-out cycle).  Hence the light curve of YLW 16A in
Figure 2 is extracted from a 2$''$.5--7$''$.5 annulus in order to avoid
the pile-up effect.  Except for this giant flare, most of the other
flares show a typical profile which is found in more evolved Class II
and IIIs: a fast rise and slow decay.

\begin{figure}
 \psbox[xsize=0.4\textwidth]{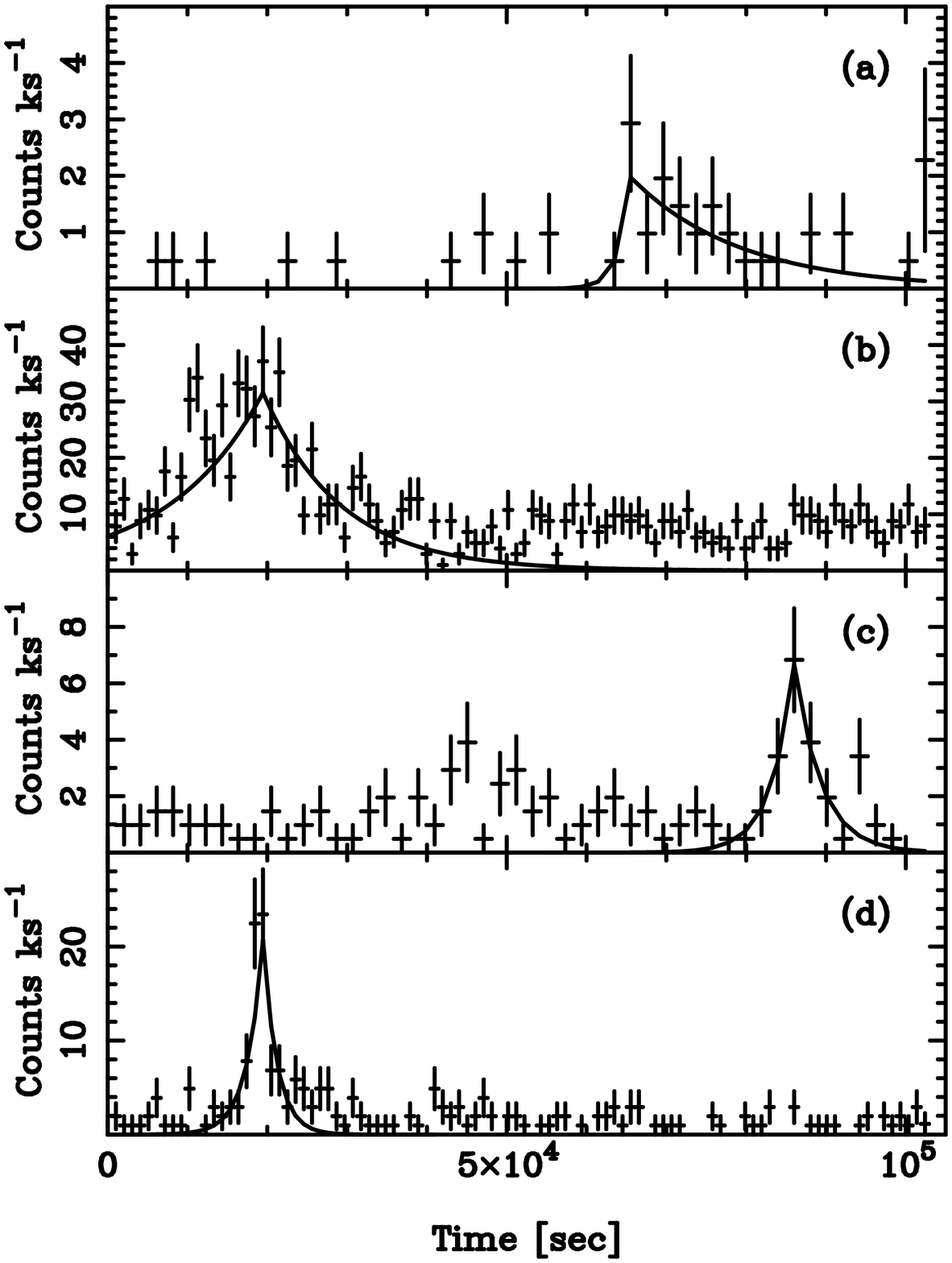}
 \psbox[xsize=0.4\textwidth]{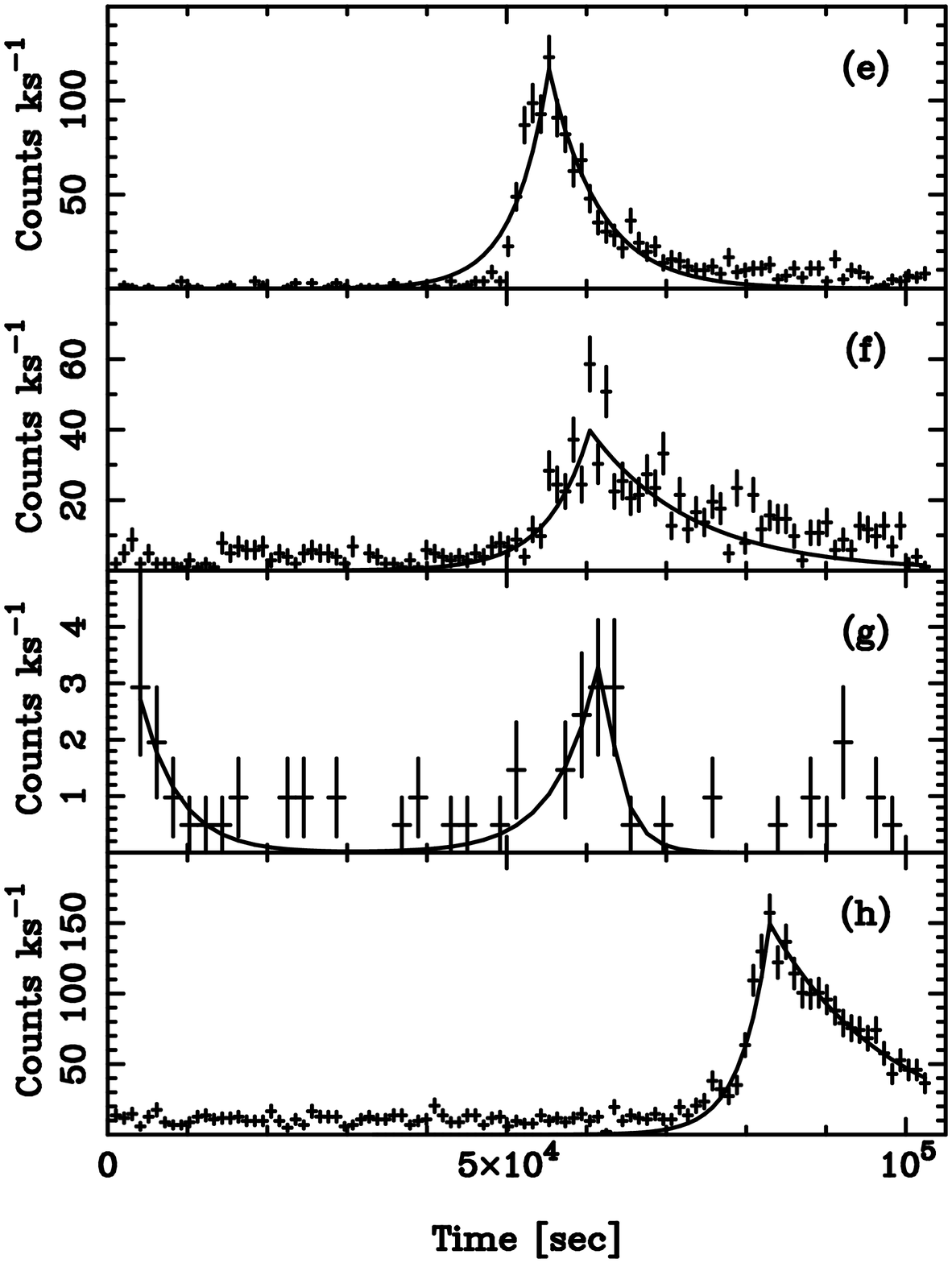}

 \psbox[xsize=0.4\textwidth]{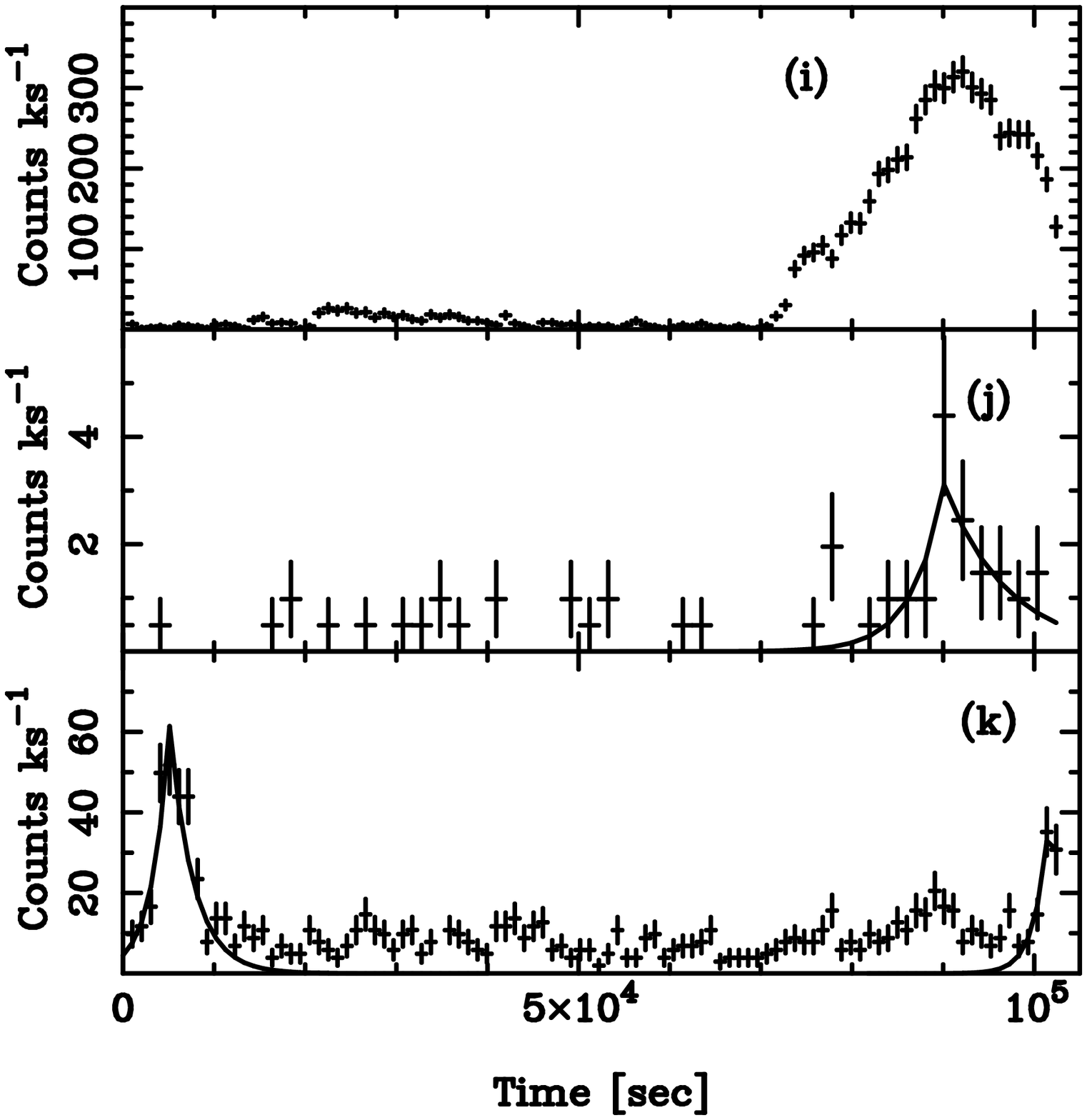}
 \caption[f2a.eps,f2b.eps,f2c.eps]{Light curves of the flaring Class
 I+I$_c$s in the total (0.7--7.0keV) X-ray band.  The horizontal axis is
 the time from the observation start (MJD = 51647.77) and the vertical
 axis is the ACIS-I count rate.  The best-fit exponential models for the
 rise and decay phases are shown by solid lines except for (i).  The light
 curve (i) is extracted from 2$''$.5--7$''$.5 radius in order to avoid
 the pile-up effect.  (a):~WL 21 (No.8); (b):~EL 29 (No.23); (c):~GY 224
 (No.24); (d):~WL 19 (No.25); (e):~YLW 10 (No.26); (f):~WL 20 (No.29);
 (g):~GY 256 (No.45); (h):~YLW 15A (No.54); (i):~YLW 16A (No.57);
 (j):~YLW 16B (No.60); (k):~YLW 45 (No.79).  \label{fig:lc_classI}}
\end{figure}

\placefigure{fig:lc_classI}

\subsection{X-ray Spectra and Luminosities}

We make and analyze the X-ray spectra of all the detected X-ray sources.
For the flare sources, we treat the spectra in quiescent and flare
phases separately. Flare phase is defined as the time from ($t_p$ $-$
$t_r$) to ($t_p$ + $t_d$), where $t_p$, $t_r$, and $t_d$ are epoch of
the flare peak and $e$-folding time for the rising and decaying phases,
respectively.  From all the source spectra, we subtract a common
background spectrum which is made from a source-free 63 arcmin$^2$
region in the ACIS-I field.  We then fit the spectra with a thin thermal
plasma model. The metal abundance is well determined only for bright
sources to be around 0.3 solar, which is consistent with the previous
{\it ASCA} results (Koyama el al. 1994; Kamata et al. 1997).  We
therefore fix the abundance to 0.3 solar unless otherwise noted, because
photon statistics are generally limited.  The free parameters are plasma
temperature ($kT$), emission measure (EM), and absorption column
($N_{\rm H}$). If the temperature is not constrained, which is often the
case for very faint sources, we fix the temperature at two
representative values of 1 keV (soft source) and 5 keV (hard source),
and estimate respective absorptions and luminosities.  These spectral
fits are generally acceptable except WL 22 (No.10), YLW 16A (No.57), and
ROXs 21 (No.40), which are treated separately in \S4.8. The best-fit
temperature, emission measure, absorption column density, flux, and
absorption-corrected luminosity in 0.5-10 keV are listed in Table 1
(columns 6--10).

Since the {\it ASCA} results are available for some of the bright ACIS
sources, we cross-check the best-fit parameters and find good agreement
between the two within the statistical errors. We are therefore
confident with the ACIS-I spectral fitting even in the rather early
phase of the $Chandra$ mission.

\section{DISCUSSION}

\subsection{X-rays from Protostars}

In the ACIS-I field, 10 Class Is and 16 Class I$_c$s have been reported
with near- to far-IR observations.  Among them, we detect X-rays from 7
Class Is and 11 Class I$_c$s above the detection limit of $\sim10^{28}$
ergs s$^{-1}$.  Thus the X-ray detection rates are as high as 70 \% in
both Class Is and Class I$_c$s.  The lack of X-rays from the other Class
I+I$_c$s could be partly due to short time variability.  For example, WL
21 (No 8) and YLW 16B (No.60) are hardly visible, except when they
exhibit an X-ray flare (see Figure 2a and 2j).  Also long-term
variability, like the 11-year cycle of solar activity, may not be
ignored.  We hence suspect that all the Class I+I$_c$s, at least those
in the $\rho$ Oph cloud, are potential X-ray sources.  Carkner, Kozak,
\& Feigelson (1998) compiled the X-ray emission from YSOs using {\it
ROSAT} PSPC and reported that the X-ray detection rate from Class
I+I$_c$s was only $\sim$10\% above the detection limit of $10^{29}$ ergs
s$^{-1}$.  Our high detection rate of Class I+I$_c$s ($\sim$70 \%) is
due to the higher sensitivity of {\it Chandra}, especially in the hard
X-ray band. In fact, if we convert the {\it Chandra} flux of the 18
Class I+I$_c$s to the PSPC counts using the {\it pimms}
software\footnote{See http://asc.harvard.edu/toolkit/pimms.jsp}, we find
only 4 sources fall above the PSPC lower limit ($\approx$ 10$^{-3}$ cts
s$^{-1}$), even in their flare phases.  This corresponds to the PSPC
detection rate of $\sim$20 \%.

To search for X-rays from an even younger phase than Class I, we check
hard X-ray enhancements from 41 starless cores (Motte, Andr\'e, \& Neri
1998) in the ACIS-I field. The 3--8 keV band photon fluxes in a 13$''$
radius circle around the 1300 $\mu$m sources are either below the 3
$\sigma$ significance level (37 sources) or are contaminated by nearby
YSOs (4 sources).

\subsection{X-rays from  Brown Dwarfs}

In our field of view, 3 BDs and 4 BD$_c$s have been reported with
infrared spectroscopy (Wilking, Greene, \& Meyer 1999; Cushing,
Tokunaga, \& Kobayashi 2000).  We detect X-rays from 1 BD (GY 310,
No.76) and 1 BD$_c$ (GY 326, No.81).  We should note that ROSAT also
found X-rays from another BD GY 202 (Neuh\"auser et al. 1999).  The ages
of these stars are estimated to be in the range of
(0.1--5)$\times$10$^{6}$ yr (Willing, Greene, \& Meyer 1999; Cushing,
Tokunaga, \& Kobayashi 2000).  Interestingly, the other non-X-ray BDs
and BD$_c$s have systematically older ages of (2--90)$\times$10$^{6}$
yr.  The X-ray properties of the BD and BD$_c$ are similar to Class
II+IIIs: relatively low absorption (0.5--2.3$\times$10$^{22}$ cm$^{-3}$)
and temperature (0.9--2.0 keV). Both have $L_{X}$/$L_{\rm bol}$ of about
3$\times$10$^{-4}$, like typical low mass YSOs.  Although the X-rays
show aperiodic variability, no flare-like event is found.

\subsection{$N_{\rm H}$ and ($J-H$)}

Using the $J$ and $H$ band magnitudes of Barsony et al. (1997), we
display the relation between the X-ray absorption column ($N_{\rm H}$)
and the IR color ($J-H$) in Figure 3.  From this figure, we see that
$N_{\rm H}$ is roughly proportional to ($J-H$).  The best-fit linear
relation is ($N_{\rm H}$/10$^{22}$cm$^{-2}$) = (1.26$\pm$0.08)\{($J-H$)
+ (0.65$\pm$0.09)\}, where the data points of No.1, 8, 23, and 77 are
excluded for the fitting (see the next paragraph). The offset of 0.65
mag corresponds to that of the zero-reddening K-M type stars (Cox 2000). 
Using the relation of $A_{\rm v}$ and infrared extinction of the $\rho$
Oph cloud (Martin \& Whittet 1990), we obtain a relation of $N_{\rm H}$
[cm$^{-2}$] = (1.59$\pm$0.40)$\times$10$^{21}$ $A_{\rm v}$, which should
be compared to $N_{\rm H}$ [cm$^{-2}$] = 1.79$\times$10$^{21}$ $A_{\rm
v}$ in the interstellar space of our Galaxy (Predehl \& Schmitt 1995).
 
Large $N_{\rm H}$ excesses are found in 3 Class I+I$_c$s: WL 12 (No.1),
WL 21 (No 8), and EL 29 (No.23).  Since the X-rays would be emitted from
the stellar surface or innermost disk, these Class I+I$_c$s should have
extra infrared photons coming from outer regions.  One possibility is
that IR photons from the stellar surface escape above or below the
accretion disk or torus, then the envelope gas scatters the IR photons
into our line of sight.  Since the scattering efficiency is larger at
shorter wavelengths, this process makes apparent ($J-H$) smaller than
that expected from $N_{\rm H}$.

An unclassified source, GY 312 (No.77) also shows large $N_{\rm H}$
excess.  The X-ray properties of this source, high $kT$, $N_{\rm H}$,
magnetic field, and $L_{X}$(flare)/$L_{X}$(quiescent) are more similar
to Class Is than Class II+IIIs (see also figures 4, 6, and 7), hence GY
312 may be a Class I.

\begin{figure}
 \psbox[xsize=0.47\textwidth]{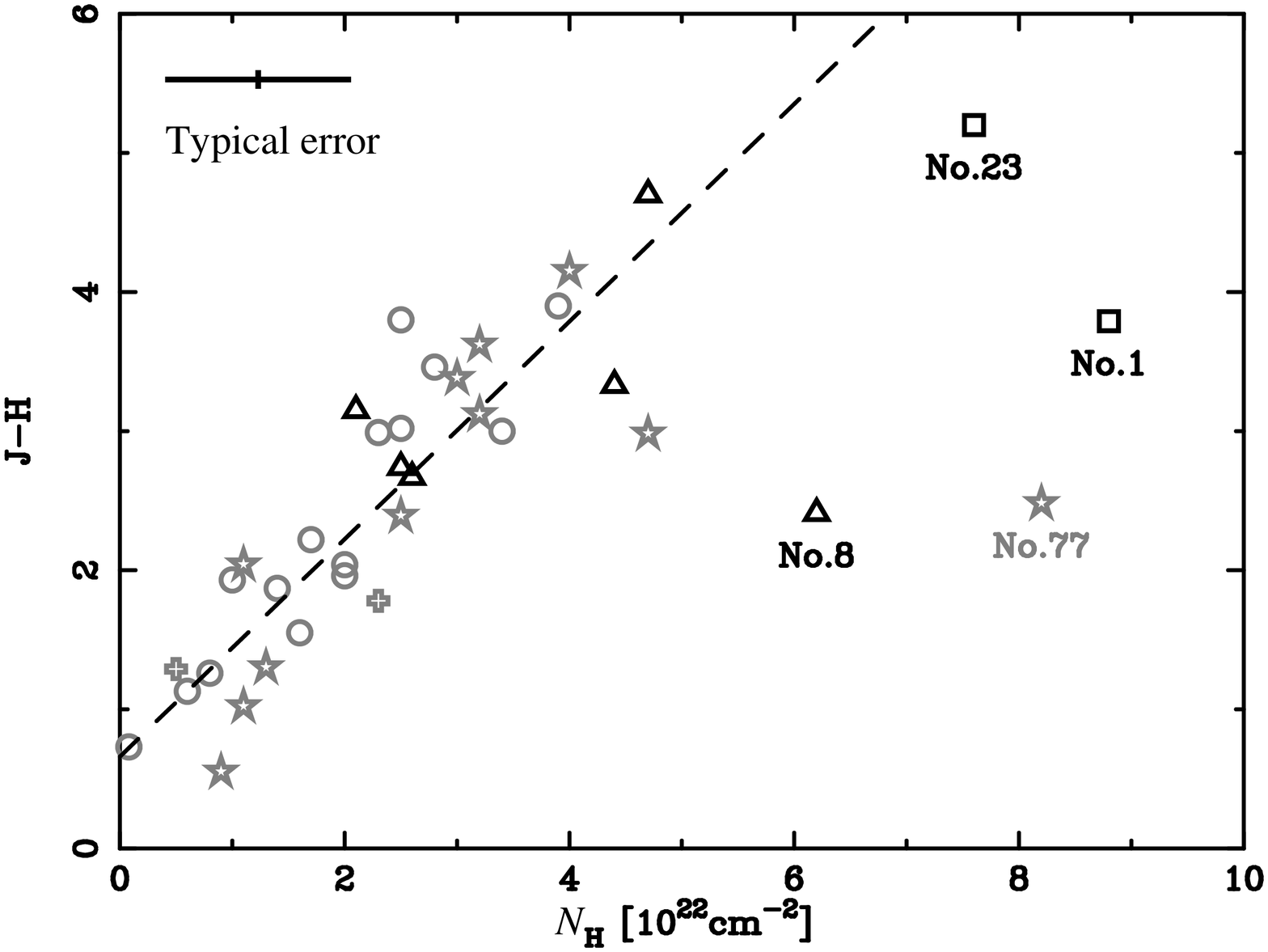}

 \caption[f3.eps]{Plot of the absorption column ($N_{\rm H}$) and the
 near-infrared color ($J-H$). Square, triangle, circle, cross, and star
 represent Class I, Class I$_c$, Class II+III, BD and BD$_c$, and
 unclassified source, respectively. The dashed line represents the
 best-fit linear model excluding the data points of No.1, 8, 23, and 77. 
 \label{fig:NH_col}}
\end{figure}

\placefigure{fig:NH_col}

\subsection{$N_{\rm H}$ and $kT$}

We display the relation between absorption ($N_{\rm H}$) and temperature
($kT$) in Figure 4.  For flare sources, data points in the quiescent and
flare phases are separately plotted.  The data are limited to bright
sources, for which both $N_{\rm H}$ and $kT$ are well constrained.  ROX
s21 (No.40) is excluded because this shows a complicated spectrum (see
\S4.8.6).  From Figure 4, we see that $N_{\rm H}$ values for class
I+I$_c$s are larger than those of Class II+IIIs.  This supports the idea
that the X-rays from Class I+I$_c$s come through denser envelopes
surrounding the protostars.

We also see that the temperatures of Class I+I$_c$s are systematically
larger than those of Class II+IIIs both in quiescent and flare
phases. This result confirms and further extends the early {\it ASCA}
results (Koyama et al. 1994; Kamata et al. 1997).

\begin{figure}
 \psbox[xsize=0.47\textwidth]{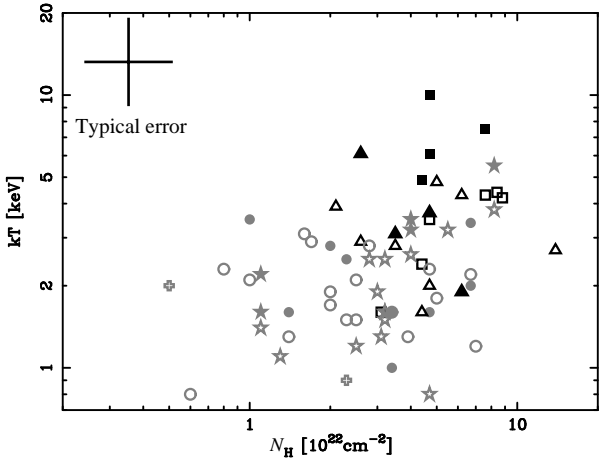}

 \caption[f4.eps]{Plot of the absorption column density ($N_{\rm H}$)
 and the plasma temperature ($kT$).  Symbols are the same as Figure
 3. For flare sources, those of quiescent and flare phases are
 separately given by open and filled symbols. Note that the flare and
 quiescent data of an unclassified source No.77 (GY 312) lie at large
 $N_{\rm H} \approx$ 8$\times$10$^{22}$ cm$^{-2}$, similar to Class Is
 (see text).  \label{fig:nh_kt}}
\end{figure}

\placefigure{fig:nh_kt}

%
%
\subsection{$L_{X}$ and $L_{\rm bol}$}

For sources whose bolometric luminosity is well determined, we display
the luminosity relation between the X-ray ($L_{X}$) and the bolometric
($L_{\rm bol}$) in Figure 5, then fit the data with a linear
function. The best-fit $L_{X}$/$L_{\rm bol}$ values in quiescent phases
are 4.4$\times$10$^{-5}$ and 4.3$\times$10$^{-4}$ for Class I+I$_c$s and
Class II+IIIs, respectively, while those in flare phases are
8.8$\times$10$^{-4}$ (Class I+I$_c$s) and 1.7$\times$10$^{-3}$ (Class
II+IIIs).  We thus confirm the $L_{X}$/$L_{\rm bol}$ $\sim$ 10$^{-4}$
relation for quiescent phases of Class II+IIIs (Feigelson et al. 1993;
Casanova et al. 1995; Grosso et al. 2000). Class I+I$_c$s, on the other
hand, show significantly lower $L_{X}$/$L_{\rm bol}$ values in both the
flare and quiescent phases than those of Class II+IIIs.  This can be
interpreted that most of $L_{\rm bol}$ of Class I+I$_c$s comes from
accreting matter rather than the stellar surface, because these objects
are in a dynamical mass-accretion phase.

We also see that the X-ray luminosities of 7 flares and 4 quiescent
phases well exceed the saturation limit ($\sim$10$^{-3}$) seen in
low-mass main-sequence stars (Fleming, Schmitt, \& Giampapa 1995).  This
supports the idea that X-ray emitting regions are larger than the
stellar surface, which is consistent with the star-disk arcade scenario
(Montmerle et al. 2000).

\begin{figure}
 \psbox[xsize=0.47\textwidth]{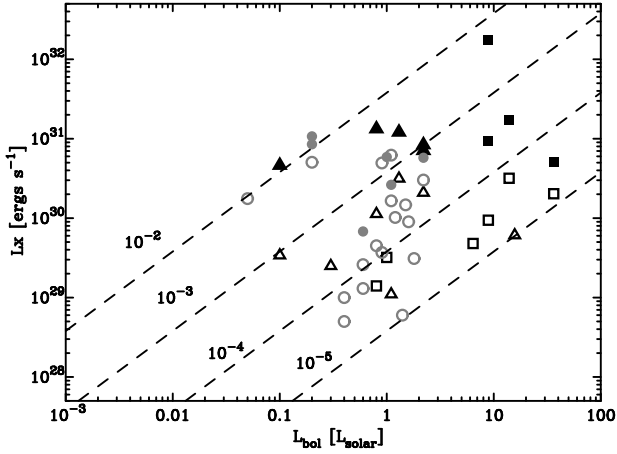}

 \caption[f5.eps]{Plot of the bolometric luminosity ($L_{\rm bol}$) and
 the X-ray luminosity ($L_{X}$).  Symbols are the same as Figure
 3. The dashed lines represent the constant $L_{X}$/$L_{\rm bol}$ ratio of
 10$^{-2}$--10$^{-5}$.  \label{fig:lx_lbol}}

\end{figure}

\placefigure{fig:lx_lbol}

\subsection{X-ray flares, $L_{X}$,  $kT$ and EM relation}
 
We have detected a number of flares, which leads us to
a systematic study.  The duty ratios of flares, 14, 12, and 9 flares from
18 Class I+I$_c$s, 20 Class II+IIIs, and 17 unclassified sources, are
nearly equal among these classes, although a hint of larger flare rate
is seen in younger phase, Class I+I$_c$s.
%
%
Regardless of the IR classes, most of the flares show typical
characteristics of stellar flares: a fast rise and slow decay in flux
and temperature.

We plot the luminosity ($L_{X}$) correlation between flare and
quiescent phases in Figure 6.  We see that $L_{X}$ in quiescent
phases is largely scattered from 2$\times$10$^{28}$ to 5$\times$10$^{30}$
ergs s$^{-1}$, irrespective of the IR classes, while
$L_{X}$(flare)/$L_{X}$(quiescent) of Class I+I$_c$s is systematically
larger than that of Class II+IIIs.

\begin{figure}
 \psbox[xsize=0.47\textwidth]{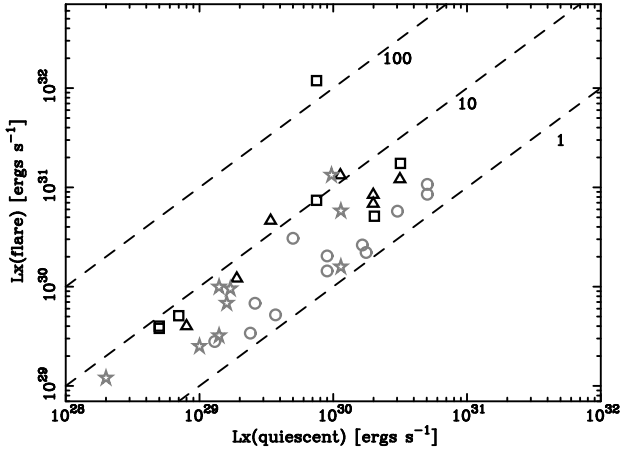}

 \caption[f6.eps]{Plot of the X-ray luminosity in the quiescent and
 flare phases. Symbols are the same as Figure 3. Dashed lines represent
 Lx(flare) = $\alpha$ Lx(quiescent) ($\alpha$ = 1, 10, 100) relations.
 An unclassified source No.77 (GY 312) lies at $\alpha$ $>$ 10, similar
 to Class Is (see text).  \label{fig:lx_f_q}}
\end{figure}

\placefigure{fig:lx_f_q}

We then make the correlation map between $kT$ and emission measure (EM)
of flare phases in Figure 7. Feldman, Laming, \& Doschek (1995) found a
universal correlation between the temperature and emission measure
ranging from solar to stellar flares.  Shibata \& Yokoyama (1999)
confirmed the observational correlation by compiling the data of
Feldman, Laming, \& Doschek (1995), solar micro flares (Shimizu 1995), a
T Tauri flare (Tsuboi et al. 1998), and a protostellar flare (Koyama et
al. 1996).  They further derived a relation of EM $\propto$
B$^{-5}T^{17/2}$, which is given by dashed lines in Figure 7 for the
magnetic fields of 15, 50, and 150 Gauss.

The data of YSOs, solar flares, and solar micro flares globally follow
the relation along the constant magnetic field (15--150 G) line, in
spite of large difference of $kT$ and EM between these objects.  In
detail, we find a hint that the magnetic fields in the flares for class
I+I$_c$s (50--150 G) are systematically larger than those for class
II+IIIs (15--50 G).  This implies that the magnetic fields in flares may
become smaller as a star evolves.

\begin{figure}
 \psbox[xsize=0.47\textwidth]{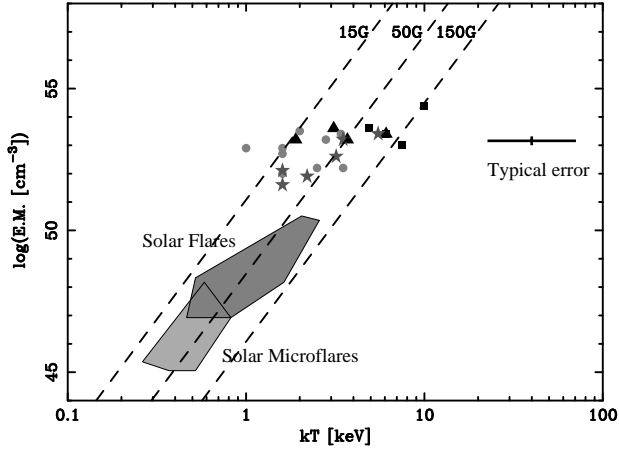}

 \caption[f7.eps]{Plot of the plasma temperature ($kT$) and the emission
 measure (EM) in the flare phase. Symbols are the same as Figure 3. The dark
 and light gray regions represent solar flares and micro flares. The dashed
 lines are the EM--$kT$ relation (EM $\propto$ B$^{-5}T^{17/2}$) by
 Shibata \& Yokoyama (1999) for B = 15, 50, 150 G.  Note that an
 unclassified source No.77 (GY 312) lies at $kT$ = 5--6 keV, similar to
 Class Is (see text).  \label{fig:kt_em}}
\end{figure}

\placefigure{fig:kt_em}

\subsection{Comments on the Unidentified Sources}

In Table 1, we find that the absorptions and temperatures of most of the
29 unidentified sources are larger than those of Class II+IIIs.  To
increase statistics, we co-add all the unidentified sources to make a
composite spectrum and fit it with a power-law model. The best-fit
photon index is 1.6$^{+0.3}_{-0.2}$, which is similar to that of the
canonical value of AGNs ($\sim$1.7).  The best-fit $N_{\rm H}$ value is
5.9$^{+1.3}_{-1.1}\times$10$^{22}$ cm$^{-2}$, nearly equal to the
average value through this dark cloud (Tachihara, Mizuno, \& Fukui
2000). Inversely, if the background sources have a mean photon index of
1.6 and are absorbed by the cloud gas of $N_{\rm H}$ =
5.9$\times$10$^{22}$ cm$^{-2}$ , the log $N$ -- log $S$ relation by
Mushotzky et al. (2000) predicts the detectable source numbers to be a
few tens.  Thus most of the unidentified sources are likely to be
background AGNs, although some fraction may be cloud members and/or
foreground sources with a small bolometric luminosity. The latter is
likely very low mass YSOs such as BDs.  No.30 would be a good candidate
for the latter case, because it exhibits a rapid flare with similar
temperature and absorption to those of Class I+I$_c$s.

\subsection{Individual Sources}

Further details of selected 6 Class I+ I$_c$ and 1 Class III sources are
given below.

\subsubsection{WL 22 (No.10)}

The X-ray spectrum of WL 22 exhibits clear line structure at $\sim$3.2
and $\sim$3.9 keV as shown in Figure 8. The line energies correspond to
He-like Ca and Ar.  We hence relax the abundances of these elements to
be free and fit with a thin thermal model. The best-fit $kT$ and $N_{\rm
H}$ are 2.7$^{+2.5}_{-1.0}$ keV and 13.9$^{+18.8}_{-5.5}\times$10$^{22}$
cm$^{-2}$, respectively (Table 1), while the abundances of Ar and Ca are
9.3$^{+15.0}_{-8.3}$ and 12.0$^{+14.1}_{-8.1}$ of solar. Thus Ca is
extremely overabundant, although the mechanism is an open question.

\begin{figure}
 \psbox[xsize=0.47\textwidth]{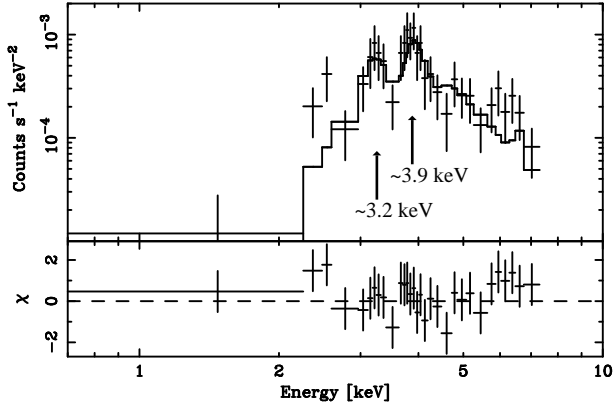} 

 \caption[f8.eps]{Spectrum of WL 22 (No.10). The upper panel shows data
 points (crosses) and the best-fit thin-thermal plasma model (solid
 line), while the lower panel shows residuals from the best-fit
 model. Line structures at $\sim$3.2 keV and $\sim$3.9 keV are due to
 He-like Ar and Ca.  \label{fig:spec_No10}}
\end{figure}

\placefigure{fig:spec_No10}

\subsubsection{EL 29 (No.23)}

EL 29 exhibits an X-ray flare with an $e$-folding time of $\sim$10 ks
(Figure 2b). The average temperature and flux in the flare are 7.5 keV
and 5.1$\times$10$^{30}$ ergs s$^{-1}$, while those in the quiescent
phase are 4.3 keV and 2.0$\times$10$^{30}$ ergs s$^{-1}$, respectively
(Table 1).  The {\it ASCA} satellite visited EL 29 twice with respective 
exposure times of 40 and 100 ks. In the first observation it
exhibited an X-ray flare similar to the present observation (Kamata et
al. 1997).  It became however quiescent at 3$\times$10$^{30}$ ergs
s$^{-1}$ in the second observation (Tsuboi et al. 2000).  Hence we infer
that EL 29 shows quiescent luminosity of $\sim$ 10$^{30}$ ergs s$^{-1}$,
and exhibits frequent X-ray flares.  

\subsubsection{WL 6 (No.44) and GY 256 (No.45)}

Kamata et al. (1997) claimed the detection of hard X-rays from WL 6 with
{\it ASCA}, although the nearby source GY 256, another Class I source,
may contaminate the spectrum. The X-ray light curve showed a sinusoidal
shape of about 20 hour period. Since they found no temperature
variation, which is usually associated with a normal flare, they claimed
that the modulation is due to stellar rotations. In the present
observation, we clearly detect hard X-rays not only from WL 6 but also
from GY 256.  Two faint flares are found from GY 256, with higher
temperature than the quiescent phase (Table 1 and Fig. 2g).  However no
sinusoidal light curve is noted from WL 6 nor from GY 256.

\subsubsection{YLW 15A (No.54)}

{\it ROSAT} HRI detected an extraordinary X-ray flare from YLW 15A
(Grosso et al. 1997).  The inferred luminosity, however, depends largely
on the assumed spectrum. We therefore re-estimate the luminosity using
our best-fit temperature (4.9 keV) and $N_{\rm H}$
(4.4$\times$10$^{22}$cm$^{-2}$) for a flare (Table 1).  The HRI count
rate at the flare peak of $\sim$17 cts ks$^{-1}$ is then converted to
the peak luminosity of 10$^{33}$ ergs s$^{-1}$, which is $\sim$6 times
brighter than that of the brightest flare seen in YLW 16A (see \S4.8.5),
but may not be extraordinary.  Subsequent {\it ASCA} observation
discovered three quasi-periodic X-ray flares with an interval of
$\sim$20 hours (Tsuboi et al. 2000).  These flares are interpreted with
a star-disk arcade conjecture (Montmerle et al. 2000), with an X-ray
emission mechanism somehow different from that of Class II+IIIs which
may have magnetic arcades within the star.  Although {\it Chandra}
confirms the hard X-ray emission and finds a typical flare from YLW 15A
(Fig. 2h), no multiple-flares are found. Thus quasi-periodic flares are
not always present but rather occasional phenomena.

\subsubsection{YLW 16A (No.57)}

The light curve of YLW 16A seems to be comprised of 2 flares; the first
is rather complex with several spike-like structures and the second is a
giant flare of unusual profile (Figure 9, left panel).  We examine the
spectral evolution by slicing the data in time as shown in Figure 9
(left panel). In phases 7--9, we extract the spectra from the same
region (a 2$''$.5--7$''$.5 radius annulus) as the light curve data,
because these phases suffer from the pile-up effect (see \S3.3). We
first fit the spectra with a thin thermal model allowing the abundance
and $N_{\rm H}$ to be free for all the phases, and find no significant
variation from phase to phase both in abundance ($\cong$ 0.3 solar) and
$N_{\rm H}$. We hence fix the abundance to be 0.3 solar and fit the
spectra assuming that $N_{\rm H}$ is consistent in all the phases. The
best-fit parameters for each time interval are shown in Table 3. As 
seen in Figure 9 (left panel), the temperature ($kT$) and emission
measure (EM) increase from phase 2 and reach peak values in phase 3,
then gradually decrease. The peak luminosity is estimated to be
1.3$\times$10$^{31}$ ergs s$^{-1}$.  These phenomena of the first flare
are similar to other flares found in Class I+I$_c$s and Class II+IIIs.

The second flare shows unusual time profiles in the flux and emission
measure (EM).  However, the temperature profile is more typical of the
normal flare; at phase 7 it increases rapidly, and nearly stays constant
or shows gradual decay. These profiles may be understood by the partial
occultation of a flare due to the stellar rotations.
%
%
We hence fit the light curve of the second flare with a model of
exponential (flare) $\times$ sinusoidal (rotation modulation) function,
and obtain the rotation period of $\sim$34 hours. 
%
%

Figure 9 (right panel) is the spectrum during the large flare. A
remarkable finding is an additional emission line feature near the 6.7
keV line of highly ionized iron. The best fit line energy is
6.4$^{+0.1}_{-0.4}$ keV, which is attributable to neutral or low ionized
iron. The most plausible origin is fluorescence from cold iron in the
circumstellar gas.  If the circumstellar gas is spherically distributed
around the X-ray source, the equivalent width of iron is estimated to be
$\approx$10 Z$_{\rm Fe}$($N_{\rm H}$/10$^{22}$ cm$^{-2}$) eV, where
Z$_{\rm Fe}$ and $N_{\rm H}$ are the abundance of iron and column
density, respectively (Inoue 1985).  Since the observed column density
of 4.7$\times$10$^{22}$ cm$^{-3}$ includes the interstellar gas, that of
circumstellar gas should be less than this value.  Using the iron
abundance of 0.3 solar, we predict the maximum equivalent width to be
$\approx$15 eV, which is significantly lower than the observed value of
$\approx$100 eV.  Hence we require non-spherical geometry; a larger
amount of gas should be present out of the line-of-sight. A possible
scenario is that YLW 16A has a disk with face-on geometry (Sekimoto et
al. 1997), and this disk is responsible for the fluorescent 6.4 keV
line.
%
%
Since we see no time-lag (reflection time scale) between the flare
on-set and the 6.4 keV iron line appearance within $\lesssim$10$^4$ sec,
the separation between the star and the reflecting region should be less
than $\lesssim$ 20 AU, consistent with the disk origin.

\begin{figure}
 \psbox[xsize=0.47\textwidth]{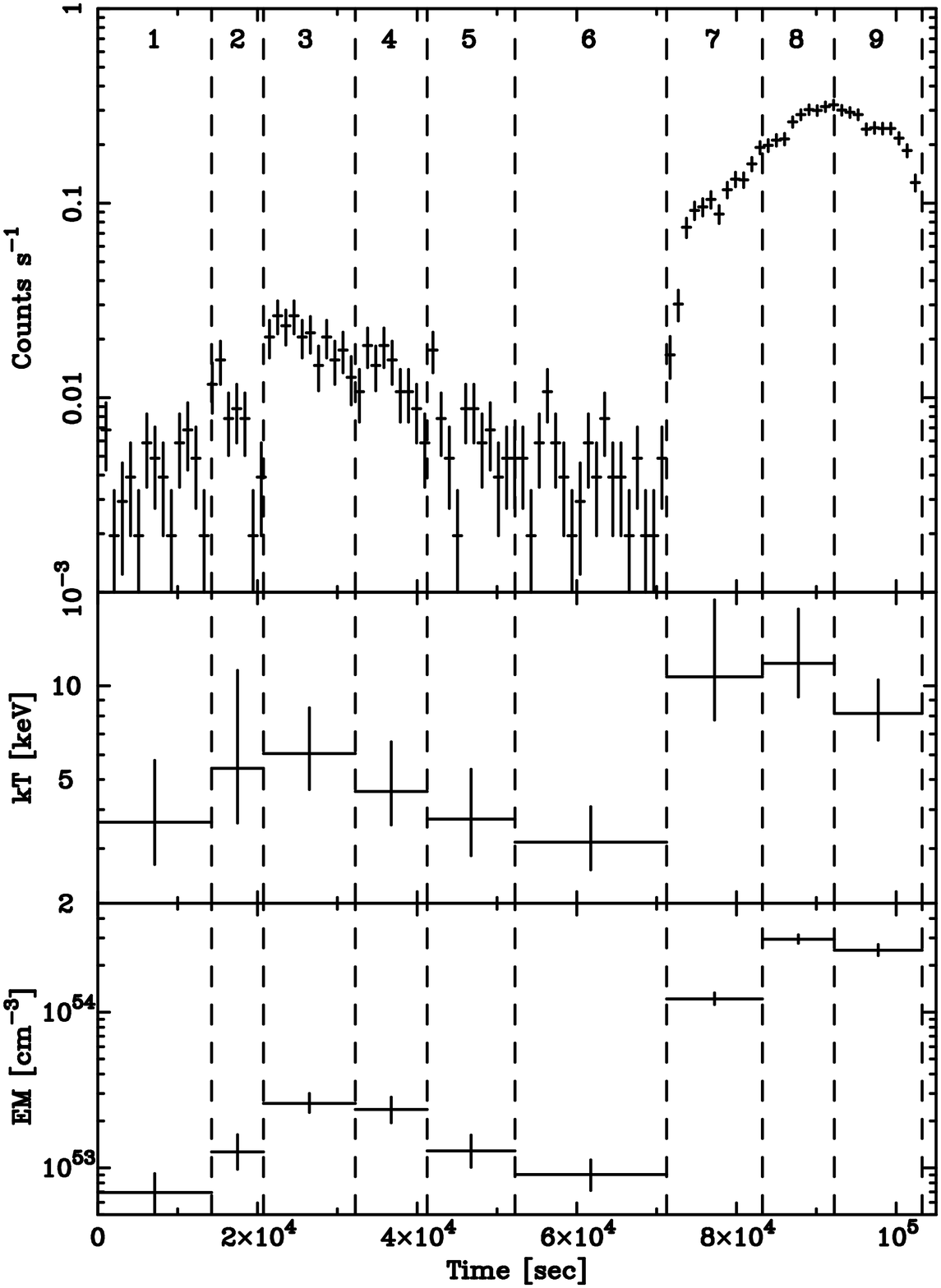}
 \psbox[xsize=0.47\textwidth]{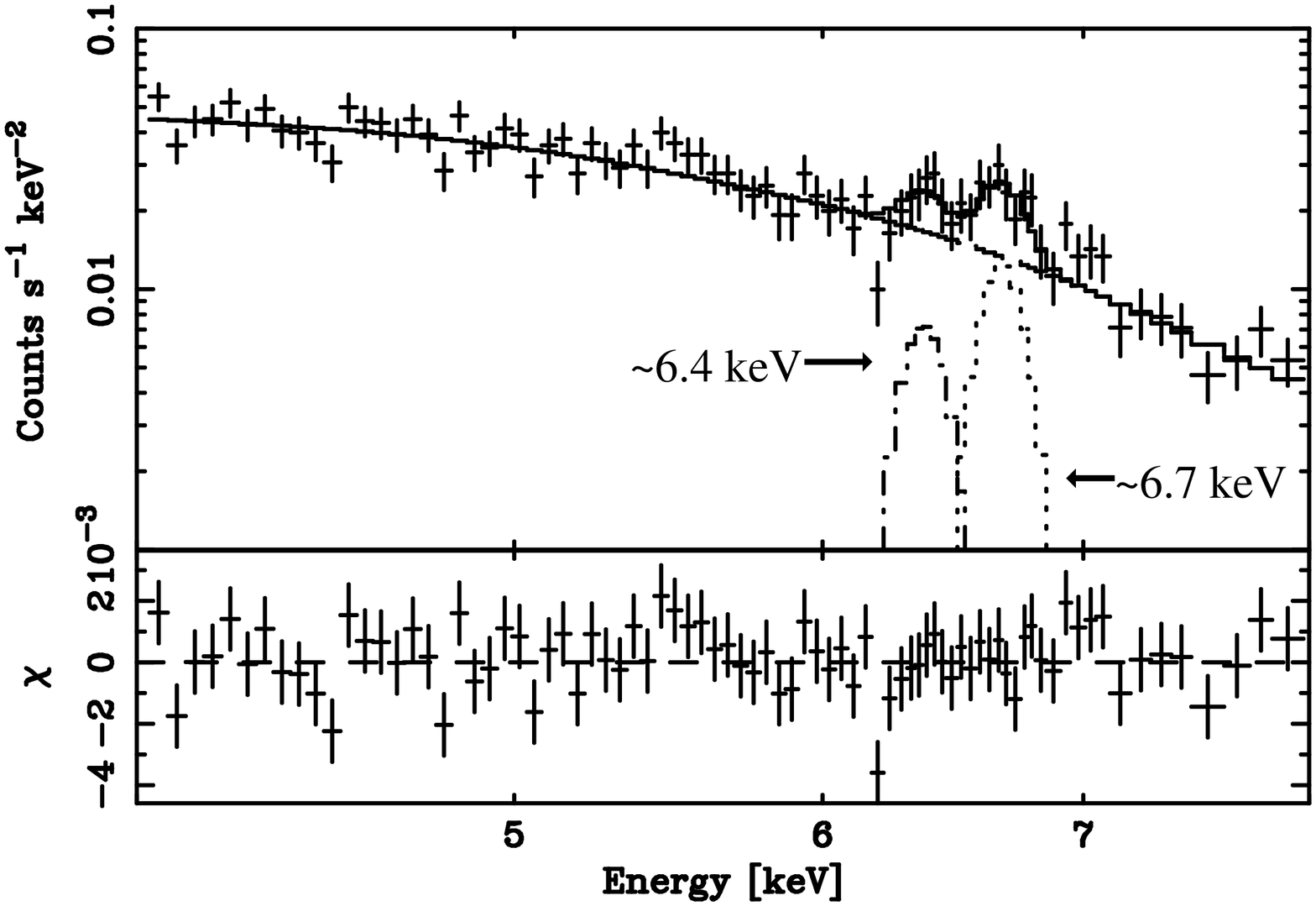} 

 \caption[f9a.eps,f9b.eps]{(left):~Light curve of YLW 16A (No.57, upper)
 and time profile of the best-fit temperature (middle) and the emission
 measure (lower).  (right):~Spectrum of YLW 16A in phases 7--9. The
 upper panel shows data points (crosses) and the best-fit model (solid
 line), while the lower panel shows residuals from the best-fit model.
 Line structures are seen at 6.4 keV and 6.7 keV.
 \label{fig:lc_para_No57}}
\end{figure}

\placefigure{fig:lc_para_No57}

\placetable{tab:ylw16a}

\subsubsection{ROX s21 (No.40)}

ROX s21, a Class III, has the brightest quiescent flux and exhibits a
flare (Figure 10, left panel).  Unlike the other sources, a single
temperature plasma fit is completely rejected. Thus we add another thin
thermal component, and find the fit to be acceptable, allowing the
abundance to be free.  The best-fit parameters are listed in Table 4.
The temperature and flux of the soft component do not change from
quiescent to flare, while those of the hard component increase during
the flare (Figure 10, right panel). In fact, if we subtract the
quiescent spectrum from that of the flare phase, we obtain a residual
5-keV plasma.  We thus suspect that ROX s21 may have fairly steady
component with a temperature of 0.84 keV and luminosity of
7$\times$10$^{29}$ ergs s$^{-1}$, and higher temperature plasma arises
from frequent flares; a larger one may be recognized as a ``flare'' (a
5-keV plasma in the present case).

\begin{figure}
 \psbox[xsize=0.47\textwidth]{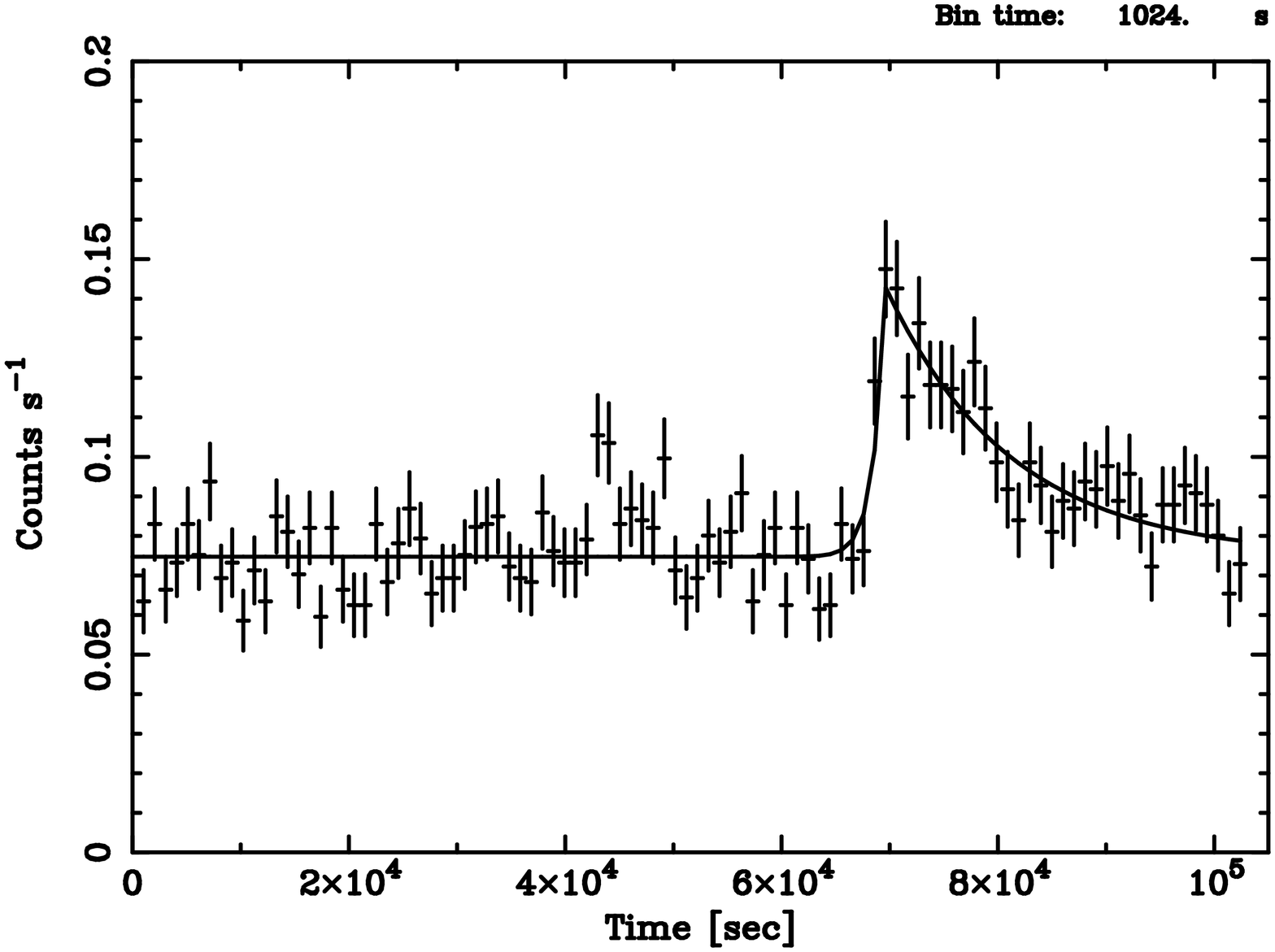}
 \psbox[xsize=0.47\textwidth]{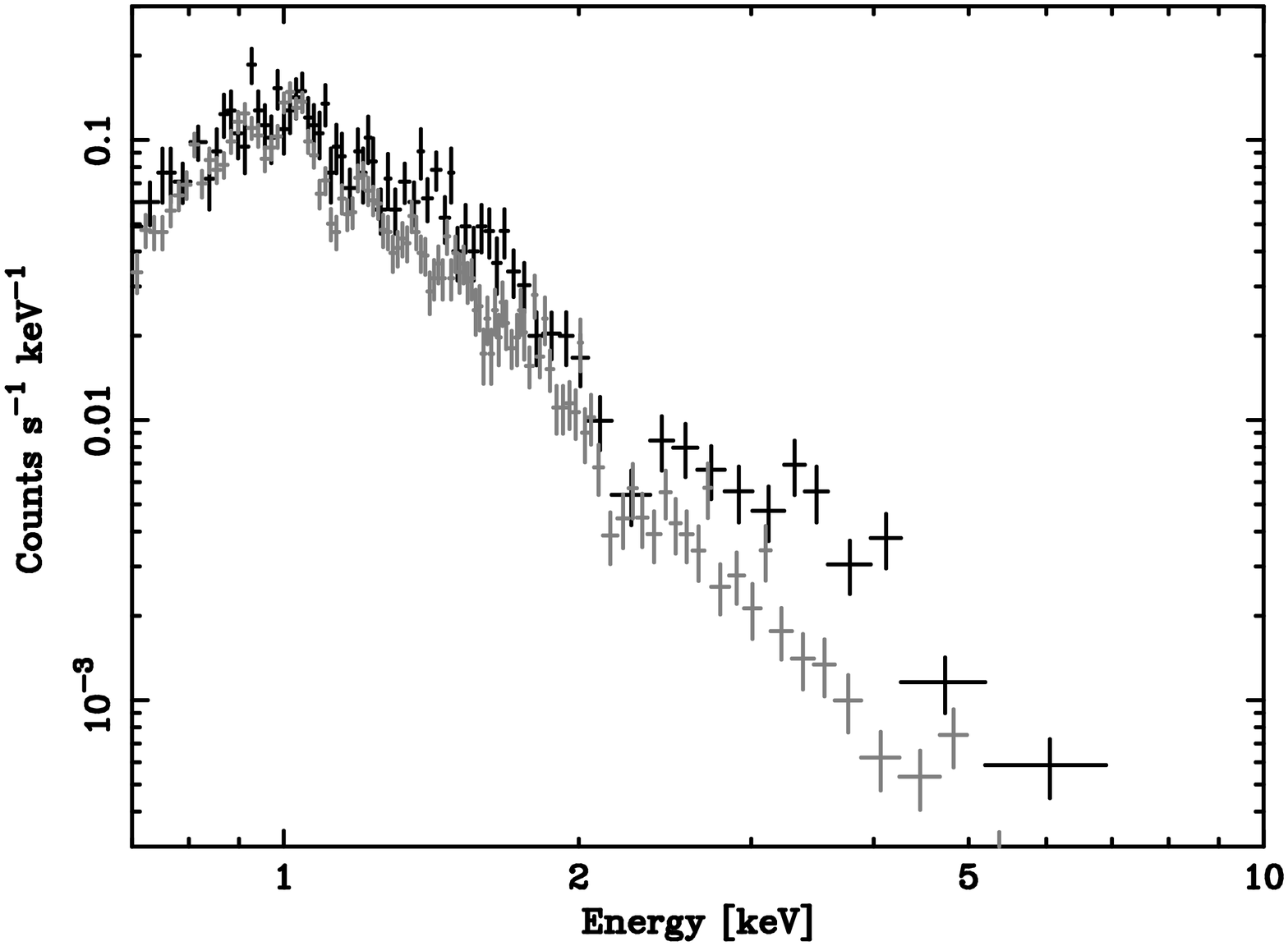}

 \caption[f10a.eps,f10b.eps]{(left):~Light curve of ROXs 21 (No.40) for
 the total X-ray band. The solid line represents the best-fit
 (exponential + constant) model. (right):~Spectra of ROXs 21 (No.40) in
 the flare (black) and quiescent (gray)
 phases. \label{fig:lc_spec_No40}}
\end{figure}

\placefigure{fig:lc_spec_No40}

\placetable{tab:roxs21}

\section{SUMMARY}

A 100-ks observation on the $\rho$ Oph cloud with $Chandra$
revealed the following:

1. We detect 87 X-ray sources from the cloud with a limiting
luminosity of $\sim$ 10$^{28}$ ergs s$^{-1}$.

2. We detect X-rays from 7  Class Is, 11
Class I$_c$s. The X-ray detection rates are both  $\sim$ 70 \%.

3. No X-rays are found from starless cores with even younger ages than
Class I.

4. We detect X-rays from 1 BD and 1 BD$_c$.  X-ray properties of these
sources are the same as Class II+IIIs of more massive stars.

5. We find 14, 12, 9, and 1 flares from 18 Class I+I$_c$s, 20 Class
II+IIIs, 17 unclassified sources, and 29 unidentified sources. Thus the
duty ratio of flares in Class I+I$_c$s is nearly equal to, or slightly
higher than that of the other classes.  Most of the flares show typical
solar-like profiles, a fast rise and slow decay, except for a giant
flare from YLW 16A.

6. Most of the X-ray spectra are well fitted with a single temperature
thin plasma model of 0.3 solar abundances.  In general, Class I+I$_c$s
have a higher temperature and absorption column than Class II+IIIs.

%
%
7. We find that Class I+I$_c$s have a systematically smaller
$L_{X}$/$L_{\rm bol}$ value than Class II+IIIs.

8. We derive the empirical relation of $N_{\rm H}$ = 1.26 E($J-H$)
10$^{22}$cm$^{-2}$ for the cloud members.

9. From the $kT$--EM relation obtained by Shibata \& Yokoyama (1999), we
infer the magnetic field to be 15--500 G, with a hint that Class
I+I$_c$s have systematically stronger magnetic fields than Class
II+IIIs.

10. Luminosity ratios between flare and quiescent phases of Class
I+I$_c$s are systematically larger than those of Class II+IIIs.

11. We find 29 unidentified sources (no IR/optical counterpart), which
are generally hard and heavily absorbed.  Most of the unidentified
sources are likely to be background AGNs.
\\

The authors express their thanks to Eric Feigelson, Gordon Garmire, and
Yoshitomo Maeda for kind hospitality to the {\it Chandra} data analysis
at PSU.  We also thank to Thierry Montmerle, Nicolas Grosso, and Leisa
Townsley for useful discussions and comments.  A part of this work is
supported by the JSPS grant of collaboration with foreign country (grant
No. 10147103).  YT is financially supported by JSPS.

\newpage
\onecolumn

%
%

\end{document}


{\scriptsize 
\begin{deluxetable}{lccccllllll}
\tablewidth{0pt}
\tablenum{1}
\tablecaption{Chandra X-ray sources in the $\rho$ Oph region \label{tab:target}
}
\tablehead{
\colhead{No.} & \colhead{R.A.} & \colhead{Dec.} & \colhead{Err} & \colhead{Cts ks$^{-1}$} &
 \colhead{$kT$} & \colhead{log(EM)} & \colhead{$N_{\rm H}$} &
 \colhead{Flux} & \colhead{$L$x} & \colhead{Phase} \nl
\colhead{(1)} & \colhead{(2)} & \colhead{(3)} & \colhead{(4)} &
 \colhead{(5)} & \colhead{(6)} & \colhead{(7)} & \colhead{(8)} &
 \colhead{(9)} &  \colhead{(10)} &  \colhead{(11)} 
}
\startdata
1 & 16:26:44.2 & $-$24:34:49.5 & 0.8 & 1.36 & 4.2(1.7--76.5) & 52.1(51.7--52.7) & 8.8(5.9--13.8) & 4.4(3.6--5.2) & 4.8(2.3--20.2) & \nl
2 & 16:26:48.6 & $-$24:28:35.3 & 2.0 & 0.59 & 1.2(0.4--4.1) & 52.3(51.4--55.0) & 7.0(3.4--19.3) & 1.0(0.7--1.3) & 5.5($>$0.7) & \nl
3 & 16:26:48.9 & $-$24:38:22.8 & 1.7 & 0.15 & 1.8(0.3--55.0) & 50.6(50.0--56.7) & 0.7($<$4.1) & 0.2(0.1--0.3) & 0.1($>$0.03) & \nl
4 & 16:26:52.0 & $-$24:30:39.5 & 1.5 & 0.26 & 2.0($>$0.5) & 51.5(50.6--53.8) & 6.4(1.7--20.4) & 0.5(0.3--0.7) & 1.0(0.1--217) & \nl
5\tablenotemark{a}  &  16:26:52.3  &  $-$24:31:33.1  & 1.4 &  0.31  &  1.0(fixed)  &  53.7(53.3--54.1)  &  59.0(41.4--83.4)  & 0.8(0.5--1.1) & 141(50.2--317) & \nl
 &   &   &   &   &  5.0(fixed)  &  51.8(51.4--52.1)  &  31.6(18.3--53.2)  & 1.3(0.8--1.8) & 2.6(1.1--5.1) & \nl
6\tablenotemark{a}  &  16:26:54.0  &  $-$24:39:24.7  & 1.6 &  0.17  &  1.0(fixed)  &  52.6(52.1--53.3)  &  21.3(12.4--52.0)  & 0.3(0.2--0.4) & 12.3(3.2--54.2) & \nl
 &   &   &   &   &  5.0(fixed)  &  51.2(50.6--51.6)  &  12.1(4.3--40.0)  & 0.6(0.3--0.9) & 0.7(0.2--2.0) & \nl
7 & 16:26:54.4 & $-$24:26:20.5 & 0.9 & 4.97 & 1.5(1.2--1.8) & 52.6(52.4--52.7) & 2.5(2.2--2.9) & 6.5(5.9--7.1) & 10.2(7.4--14.4) & \nl
8  &  16:26:57.3  &  $-$24:35:38.7  & 0.8 &  0.33  &  2.9($>$0.6)  &  52.0(51.4--54.4)  &  6.2(2.3--19.5)  & 3.4(1.8--5.0) & 4.0(1.0--880) &  flare \nl
 &   &   &   &   &  1.4($>$0.4)  &  51.5(50.5--53.7)  &   & 0.3(0.1--0.5) & 0.8(0.08--120) &  quiescent \nl
9 & 16:26:59.1 & $-$24:35:57.2 & 0.5 & 0.37 & 0.5(0.2--1.6) & 52.1(50.9--54.5) & 2.5(1.6--4.1) & 0.2(0.1--0.3) & 3.6(0.2--964) & \nl
10\tablenotemark{c} & 16:26:59.2 & $-$24:34:58.6 & 0.6 & 1.51 & 2.7(1.7--5.2) & 52.3(51.9--52.8) & 13.9(8.4--32.7) & 3.5(2.9--4.1) & 6.1(2.4--19.3) & \nl
11 & 16:27:00.0 & $-$24:26:40.4 & 1.0 & 2.35 & 7.8($>$2.6) & 52.1(52.0--52.6) & 8.1(5.7--10.5) & 8.6(7.5--9.7) & 6.9(4.8--21.4) & \nl
12\tablenotemark{a}  &  16:27:03.1  &  $-$24:26:12.9  & 2.7 &  0.07  &  1.0(fixed)  &  52.4(51.1--53.0)  &  17.2(2.6--40.4)  & 0.3(0.1--0.5) & 7.8(0.4--29.6) & \nl
 &   &   &   &   &  5.0(fixed)  &  50.7\tablenotemark{g}  &  3.8\tablenotemark{g}  & 0.3(0.1--0.5) & 0.2\tablenotemark{g} & \nl
13 & 16:27:04.1 & $-$24:28:29.3 & 0.8 & 0.96 & 3.9(2.0--13.0) & 51.4(51.1--51.7) & 2.1(1.2--3.2) & 1.6(1.3--1.9) & 1.1(0.6--2.3) & \nl
14  &  16:27:04.5  &  $-$24:27:14.9  & 1.1 &  1.07  &  1.6(0.8--6.7)  &  52.1(51.5--52.7)  &  3.2(2.1--4.8)  & 2.0(1.0--3.0) & 3.2(0.8--15.6) &  flare1 \nl
 &   &   &   &   &  3.3($>$1.3)  &  52.4(52.1--52.9)  &   & 11.9(7.8--16.0) & 9.9(4.6--32.4) &  flare2 \nl
 &   &   &   &   &  1.6(0.9--2.5)  &  51.7(51.3--52.1)  &   & 0.8(0.6--1.0) & 1.4(0.5--3.5) &  quiescent \nl
15 & 16:27:05.3 & $-$24:41:15.0 & 1.6 & 0.40 & 19.4($>$1.3) & 51.1(50.9--52.6) & 5.6(2.8--21.3) & 1.3(0.9--1.7) & 0.8(0.5--22.2) & \nl
16 & 16:27:05.9 & $-$24:26:17.5 & 2.0 & 0.46 & 1.3(0.7--9.2) & 51.7(50.9--52.5) & 3.1(1.3--5.2) & 0.5(0.4--0.6) & 1.3(0.2--7.9) & \nl
17  &  16:27:06.6  &  $-$24:41:49.8  & 1.3 &  0.66  &  1.6(0.9--5.8)  &  51.6(51.1--52.1)  &  1.1(0.2--2.2)  & 1.3(0.9--1.7) & 1.2(0.3--3.8) &  flare \nl
 &   &   &   &   &  4.6($>$0.6)  &  50.7(50.4--51.5)  &   & 0.5(0.3--0.7) & 0.2(0.1--1.3) &  quiescent \nl
18 & 16:27:06.8 & $-$24:38:15.2 & 0.5 & 1.10 & 1.6(1.0--2.7) & 52.1(51.6--52.6) & 4.4(3.7--6.7) & 1.5(1.2--1.8) & 3.1(1.1--11.9) & \nl
19\tablenotemark{a}  &  16:27:07.7  &  $-$24:41:04.1  & 1.8 &  0.11  &  1.0(fixed)  &  52.6(51.8--53.1)  &  22.2(8.8--43.7)  & 0.3(0.1--0.5) & 11.6(1.6--37.1) & \nl
 &   &   &   &   &  5.0(fixed)  &  51.0(50.1--51.3)  &  7.6(1.9--25.0)  & 0.4(0.2--0.6) & 0.4(0.07--1.0) & \nl
20\tablenotemark{a}  &  16:27:08.4  &  $-$24:39:39.7  & 1.4 &  0.12  &  1.0(fixed)  &  53.6(52.5--54.5)  &  62.9(25.2--132.0)  & 0.5(0.2--0.8) & 102(8.5--942) & \nl
 &   &   &   &   &  5.0(fixed)  &  51.7(50.6--52.6)  &  36.8(4.1--102.3)  & 1.0(0.4--1.6) & 2.3(0.2--18.4) & \nl
21 & 16:27:09.1 & $-$24:34:08.2 & 0.3 & 0.73 & 2.9(1.4--12.9) & 51.2(50.9--51.6) & 1.7(1.3--2.8) & 0.8(0.6--1.0) & 0.6(0.3--1.6) & \nl
22 & 16:27:09.3 & $-$24:40:22.1 & 0.8 & 0.42 & 0.8(0.4--1.5) & 52.2(51.5--53.4) & 4.7(2.9--7.8) & 0.5(0.3--0.7) & 4.8(0.9--71.8) & \nl
23  &  16:27:09.4  &  $-$24:37:19.0  & 0.1 &  11.1  &  7.5(4.9--16.2)  &  53.0(52.9--53.1)  &  7.6(6.8--8.5)  & 64.4(58.5--70.3) & 51.2(41.7--67.9) &  flare \nl
 &   &   &   &   &  4.3(3.2--6.0)  &  52.7(52.6--52.8)  &   & 20.2(18.6--21.8) & 20.3(15.6--28.1) &  quiescent \nl
24\tablenotemark{d}  &  16:27:11.2  &  $-$24:40:46.6  & 0.8 &  1.18  &  $>$1.9  & 52.3(52.1--52.5) & 5.2(3.6--7.4) & 49.2(29.1--69.3) & 22.7(13.0--35.1) &  flare \nl
 &   &   &   &   &  4.5($>$2.0) & 51.7(51.4--52.1) & & 2.8(2.2--3.4) & 2.5(1.4--6.9) &  quiescent \nl
25  &  16:27:11.7  &  $-$24:38:32.3  & 0.3 &  2.13  &  1.9(1.2--3.6)  &  53.2(52.8--53.7)  &  6.2(4.6--8.5)  & 24.5(18.0--31.0) & 46.2(19.2--139.6) &  flare \nl
 &   &   &   &   &  4.3(2.3--10.7)  &  51.9(51.7--52.3)  &   & 3.7(3.1--4.3) & 3.4(1.9--7.7) &  quiescent \nl
26  &  16:27:15.4  &  $-$24:26:39.5  & 0.4 &  13.6  &  3.1(2.6--3.9)  &  53.6(53.5--53.7)  &  3.5(3.2--3.9)  & 148(138--158) & 133(109--168) &  flare \nl
 &   &   &   &   &  2.8(2.3--3.4)  &  52.5(52.4--52.6)  &   & 11.7(10.7--12.7) & 11.3(9.2--14.6) &  quiescent \nl
27 & 16:27:15.5 & $-$24:30:53.7 & 0.7 & 0.32 & 1.8(0.5--27.4) & 51.6(50.8--54.0) & 5.0(2.2--17.0) & 0.6(0.4--0.8) & 1.0(0.2--267) & \nl
28 & 16:27:15.6 & $-$24:33:01.6 & 0.3 & 0.40 & 6.4($>$0.4) & 51.2(50.9--55.5) & 6.3(2.9--52.0) & 1.0(0.7--1.3) & 0.8($>$0.4) & \nl
29  &  16:27:15.9  &  $-$24:38:43.6  & 0.2  &  10.4  &  6.1(3.9--11.3)  &  53.4(53.3--53.5)  &  2.6(2.3--3.0)  & 199(181--217) & 121(101--157) &  flare \nl
 &   &   &   &   &  2.9(2.3--3.6)  &  52.9(52.8--53.1)  &   & 37.3(34.2--40.4) & 31.5(24.9--41.2) &  quiescent \nl
30\tablenotemark{d}  &  16:27:16.4  &  $-$24:31:14.7  & 0.4 &  1.21  &  $>$3.0  & 52.0(51.8--52.2) & 6.9(4.9--10.0) & 24.9(15.8--34.0) & 11.9(7.4--17.6) &  flare \nl
 &   &   &   &   &  4.2($>$2.0) & 51.8(51.5--52.2) &  & 3.0(2.4--3.6) & 3.1(1.6--9.1) &  quiescent \nl
31\tablenotemark{d} & 16:27:17.0 & $-$24:36:21.5 & 0.5 & 0.08 & $>$0.4  & 49.9(49.0--50.4) & 1.0(0.0--16.1) & 0.2(0.1--0.3) & 0.1(0.01--0.3) & \nl
32\tablenotemark{d} & 16:27:17.5 & $-$24:36:27.0 & 0.5 & 0.18 & $>$0.6 & 51.5(50.8--51.9) & 23.6(4.4--56.1) & 5.3(2.8--7.8) & 3.1(0.7--8.4) & \nl
33\tablenotemark{d}  &  16:27:17.6  &  $-$24:28:56.3  & 0.5 &  1.15  &  $>$1.7 &  51.8(51.6--52.0) & 5.8(4.0--8.8) & 13.9(8,4--19.4) & 6.5(4.1--9.8) &  flare \nl
 &   &   &   &   &  7.2($>$2.4) & 51.7(51.5--52.1) &  & 4.7(3.7--5.7) & 3.2(2.1--9.6) &  quiescent \nl
34  &  16:27:18.2  &  $-$24:28:52.4  & 0.2 &  18.0  &  2.0(1.6--2.7)  &  53.5(53.3--53.6)  & 6.7(6.2--7.2) & 44.5(37.7--51.3) & 85.2(56.3--128.6) &  flare1 \nl
 &   &   &   &   &  3.4(2.6--4.7)  &  53.4(53.3--53.6)  &   & 95.6(86.5--105) & 107(79.8--144) &  flare2 \nl
 &   &   &   &   &  2.2(1.9--2.5)  &  53.2(53.1--53.3)  &   & 30.4(28.7--32.1) & 50.5(40.4--64.5) &  quiescent \nl
35\tablenotemark{a}  &  16:27:18.4  &  $-$24:39:14.4  & 1.0 &  0.11  &  1.0(fixed)  &  52.3(51.7--52.6)  &  14.7(8.5--27.3)  & 0.2(0.1--0.3) & 5.1(1.4--11.8) & \nl
 &   &   &   &   &  5.0(fixed)  &  50.7(50.1--51.1)  &  6.2(2.0--20.8)  & 0.3(0.1--0.5) & 0.3(0.07--0.6) & \nl
36  &  16:27:18.5  &  $-$24:29:05.7  & 0.3 &  5.49  &  1.0(0.9--1.2)  &  52.9(52.7--53.1)  &  3.4(2.9--3.8)  & 5.1(4.5--5.7) & 20.4(12.7--33.3) &  flare1 \nl
 &   &   &   &   &  1.6(1.3--2.0)  &  52.7(52.6--52.9)  &   & 8.2(7.2--9.2) & 14.4(10.2--20.7) &  flare2 \nl
 &   &   &   &   &  1.6(0.9--5.0)  &  52.5(52.1--52.9)  &   & 5.5(3.0--8.0) & 9.0(3.3--19.9) &  quiescent \nl
37\tablenotemark{a}  &  16:27:19.0  &  $-$24:33:44.1  & 0.5 &  0.08  &  1.0(fixed)  &  51.9(51.1--52.3)  &  11.5(5.4--26.2)  & 0.1(0.02--0.2) & 2.0(0.4--6.1) & \nl
 &   &   &   &   &  5.0(fixed)  &  50.5(49.8--50.9)  &  5.0(1.3--17.6)  & 0.2(0.1--0.3) & 0.2(0.04--0.5) & \nl
38  &  16:27:19.2  &  $-$24:28:43.8  & 0.3 &  7.14  &  1.6(1.4--1.9)  &  52.9(52.7--53.1)  &  4.7(4.2--5.3)  & 10.5(9.6--11.4) & 22.1(15.2--31.8) &  flare \nl
 &   &   &   &   &  2.3(1.8--3.1)  &  52.7(52.6--52.9)  &   & 13.1(11.3--14.9) & 17.6(12.6--24.6) &  quiescent \nl
39\tablenotemark{d} & 16:27:19.3 & $-$24:31:30.7 & 0.4 & 0.53 & $>$1.7 & 51.4(51.2--51.6) & 12.7(7.9--21.5) & 5.6(4.1--7.1) & 2.9(1.8--4.6) & \nl
40 & 16:27:19.5 & $-$24:41:40.6 & 0.1 & 83.8 & \multicolumn{6}{c}{--------------- see table \ref{tab:roxs21} ---------------} \nl
41\tablenotemark{b} & 16:27:19.9 & $-$24:35:31.9 & 0.8 & 0.06 & \nodata & \nodata & \nodata & \nodata & \nodata & \nl
42 & 16:27:20.0 & $-$24:30:13.6 & 0.6 & 0.14 & 5.6($>$0.3) & 51.1(50.5--56.5) & 13.8(3.5--82.2) & 0.5(0.2--0.8) & 0.6($>$0.1) & \nl
43 & 16:27:21.4 & $-$24:41:43.2 & 0.9 & 1.39 & 1.3(0.9--2.0) & 52.2(51.8--52.6) & 3.9(2.7--5.2) & 1.7(1.4--2.0) & 4.5(1.9--11.5) & \nl
44 & 16:27:21.8 & $-$24:29:53.4 & 0.5 & 1.21 & 4.4(2.0--11.8) & 51.9(51.6--52.4) & 8.4(6.1--12.7) & 3.1(2.5--3.7) & 3.2(1.7--10.1) & \nl
45  &  16:27:22.0  &  $-$24:29:39.6  & 0.8 &  0.56  &  4.1($>$0.8)  &  52.0(51.5--53.0)  &  3.1(1.3--5.8)  & 5.4(2.0--8.8) & 4.0(1.5--46.7) &  flare1 \nl
 &   &   &   &   &  2.9($>$0.7)  &  52.0(51.6--53.1)  &   & 4.3(2.4--6.2) & 3.8(1.3--47.6) &  flare2 \nl
 &   &   &   &   &  1.6(0.5--40.3)  &  51.3(50.5--52.5)  &   & 0.3(0.2--0.4) & 0.5(0.08--9.0) &  quiescent \nl
46\tablenotemark{b} & 16:27:22.6 & $-$24:34:15.1 & 0.4 & 0.04 & \nodata & \nodata & \nodata & \nodata & \nodata & \nl
47 & 16:27:23.6 & $-$24:31:57.0 & 0.6 & 0.17 & 1.2($>$0.2) & 53.2(50.9--55.6) & 67.6(15.2--156.7) & 0.6(0.3--0.9) & 42.6($>$0.2) & \nl
48  &  16:27:24.2  &  $-$24:29:29.4  & 0.7 &  0.52  &  3.6\tablenotemark{g}  &  52.4($<$55.7)  &  1.8(0.8--4.4)  & 14.5($<$35.0) & 9.5($>$0.0) &  flare \nl
 &   &   &   &   &  8.1($>$1.5)  &  51.5(51.3--52.2)  &   & 3.3(2.3--4.3) & 1.7(1.0--8.9) &  quiescent \nl
49\tablenotemark{a}  &  16:27:24.4  &  $-$24:41:49.4  & 1.9 &  0.11  &  1.0(fixed)  &  51.2(49.9--51.7)  &  3.0(0.7--7.9)  & 0.1(0.04--0.2) & 0.5(0.02--1.5) & \nl
 &   &   &   &   &  5.0(fixed)  &  50.2\tablenotemark{g}  &  1.0\tablenotemark{g}  & 0.1(0.04--0.2) & 0.1\tablenotemark{g} & \nl
50 & 16:27:24.6 & $-$24:29:35.3 & 0.7 & 0.65 & 2.5(1.1--10.7) & 51.7(51.2--52.5) & 2.8(1.3--5.2) & 1.9(1.4--2.4) & 1.8(0.6--9.4) & \nl
51\tablenotemark{b} & 16:27:25.0 & $-$24:34:35.3 & 0.6 & 0.04 & \nodata & \nodata & \nodata & \nodata & \nodata & \nl
52\tablenotemark{e}  & 16:27:26.4 & $-$24:25:55.6 & 2.1 & 0.14 & 0.1(0.04--0.3) & 52.9(50.1--58.7) & 0.9(0.3--1.9) & 0.1(0.05--0.2) & 16.6($>$0.02) & \nl
53\tablenotemark{d}  &  16:27:26.5  &  $-$24:39:23.1  & 0.3 &  5.94  &  $>$6.9 & 52.7(52.7--52.8) & 2.8(2.5--3.2) & 136(120--152) & 58.9(51.9--66.0) &  flare \nl
 &   &   &   &   &  2.7(2.2--3.6) & 52.2(52.1--52.3) & & 6.0(5.3--6.7) & 6.6(5.0--8.6) &  quiescent \nl
54  &  16:27:27.0  &  $-$24:40:50.9  & 0.2 &  29.2  &  4.9(4.2--6.2)  &  53.6(53.5--53.6)  &  4.4(4.1--4.7)  & 228(217--239) & 174(151--198) &  flare \nl
 &   &   &   &   &  2.4(2.1--2.8)  &  53.0(52.9--53.1)  &   & 26.5(25.0--28.0) & 31.8(26.4--38.1) &  quiescent \nl
55  &  16:27:27.1  &  $-$24:32:17.6  & 0.1 &  10.8  &  3.2(2.3--5.9)  &  52.6(52.5--52.8)  &  4.0(3.5--4.5)  & 17.1(14.2--20.0) & 15.8(10.8--22.8) &  flare1 \nl
 &   &   &   &   &  3.5(2.9--4.8)  &  53.2(53.0--53.2)  &   & 66.0(60.2--71.8) & 57.8(43.6--65.1) &  flare2 \nl
 &   &   &   &   &  2.6(2.1--3.3)  &  52.5(52.4--52.7)  &   & 10.5(9.5--11.5) & 11.4(8.2--15.4) &  quiescent \nl
56  &  16:27:27.4  &  $-$24:31:16.4  & 0.2 &  4.86  &  3.5(2.4--5.0)  &  52.2(52.1--52.4)  &  1.0(0.8--1.3)  & 12.0(10.0--14.0) & 6.8(5.1--9.6) &  flare \nl
 &   &   &   &   &  2.1(1.6--2.8)  &  51.9(51.8--52.1)  &   & 3.7(3.3--4.1) & 2.6(2.0--3.5) &  quiescent \nl
57 & 16:27:28.0 & $-$24:39:33.5 & 0.1 & 188\tablenotemark{f} & \multicolumn{6}{c}{--------------- see table \ref{tab:ylw16a} ---------------} \nl
58 & 16:27:28.5 & $-$24:29:42.0 & 0.9 & 0.24 & 5.9($>$0.8) & 51.1(50.7--53.1) & 6.3(2.7--22.4) & 0.8(0.5--1.1) & 0.6(0.3--54.0) & \nl
59  &  16:27:28.5  &  $-$24:27:20.6  & 0.7 &  3.06  &  2.5(1.7--4.2)  &  52.2(52.0--52.4)  &  2.3(1.8--2.8)  & 5.9(4.7--7.1) & 5.2(3.3--8.0) &  flare \nl
 &   &   &   &   &  1.5(1.2--1.9)  &  52.1(51.9--52.3)  &   & 2.6(2.2--3.0) & 3.7(2.3--5.7) &  quiescent \nl
60  &  16:27:29.4  &  $-$24:39:15.9  & 0.9 &  0.41  &  3.0($>$0.6) & 52.2(51.6--54.5) & 11.1(4.0--18.0) & 4.1(2.1--6.2) & 7.5(1.8--1464) &  flare \nl
\phantom{60}\tablenotemark{d} &   &   &   &   &  $>$0.5  & 51.1(50.7--51.4) & & 2.8(1.6--4.0) & 1.4(0.6--2.8) &  quiescent \nl
61\tablenotemark{d}  &  16:27:30.0  &  $-$24:33:36.7  & 0.3 &  0.98  &  $>$1.6  &  52.1(51.8--52.3)  &  5.5(3.7--8.2)  & 27.9(15.4--40.4) & 12.2(7.1--18.9) &  flare \nl
 &   &   &   &   &  3.2(1.5--23.2)  &  51.6(51.3--52.2)  &   & 1.6(1.2--2.0) & 1.9(0.9--6.5) &  quiescent \nl
62\tablenotemark{d}   & 16:27:30.1 & $-$24:38:03.4 & 0.9 & 0.11 & $>$0.5 & 50.3(49.3--50.8) & 2.1($<$21.1) & 0.5(0.2--0.8) & 0.2(0.02--0.6) & \nl
63 & 16:27:30.1 & $-$24:27:43.5 & 0.8 & 2.31 & 2.1(1.3--3.6) & 52.0(51.7--52.4) & 2.5(1.8--3.8) & 3.0(2.6--3.4) & 3.1(1.6--8.2) & \nl
64  &  16:27:31.1  &  $-$24:34:03.3  & 0.3 &  2.00  &  2.2(1.1--6.4)  &  51.9(51.7--52.2)  &  1.1(0.8--1.7)  & 3.6(2.7--4.5) & 2.5(1.4--5.1) &  flare \nl
 &   &   &   &   &  1.4(1.0--2.2)  &  51.6(51.3--51.9)  &   & 1.0(0.8--1.2) & 1.0(0.6--2.4) &  quiescent \nl
65 & 16:27:31.3 & $-$24:31:10.1 & 0.5 & 0.73 & 7.0($>$2.1) & 51.4(51.2--52.0) & 5.6(3.4--10.3) & 1.9(1.5--2.3) & 1.4(0.8--5.8) & \nl
66 & 16:27:32.7 & $-$24:33:24.2 & 0.2 & 3.57 & 2.5(1.8--3.7) & 52.2(52.1--52.5) & 3.2(2.6--3.9) & 5.7(5.1--6.3) & 5.8(3.7--9.6) & \nl
67 & 16:27:32.9 & $-$24:32:35.0 & 0.3 & 1.85 & 1.9(1.3--2.6) & 52.0(51.7--52.3) & 3.0(2.3--4.0) & 2.2(1.9--2.5) & 2.8(1.6--5.8) & \nl
68 & 16:27:33.0 & $-$24:31:48.2 & 1.4 & 0.14 & 1.9($>$0.1) & 51.2(50.0--62.5) & 7.4(1.0--27.5) & 0.2(0.1--0.3) & 0.5($>$0.03) & \nl
69  &  16:27:33.1  &  $-$24:41:15.3  & 0.2 &  31.3  &  2.8(2.5--3.3)  &  53.2(53.1--53.3)  &  2.0(1.9--2.1)  & 74.6(70.5--78.7) & 57.6(49.4--64.2) &  flare \nl
 &   &   &   &   &  1.9(1.8--2.2)  &  53.0(52.9--53.1)  &   & 29.6(28.2--31.0) & 30.2(25.5--33.1) &  quiescent \nl
70 & 16:27:35.3 & $-$24:38:33.7 & 0.7 & 1.03 & 0.8(0.6--1.1) & 51.2(51.0--51.5) & 0.6(0.4--0.9) & 0.4(0.3--0.5) & 0.5(0.3--0.9) & \nl
71\tablenotemark{d}   & 16:27:36.0 & $-$24:32:18.4 & 1.0 & 0.26 & $>$3.0 & 50.7(50.4--51.0) & 2.1(0.8--6.7) & 1.4(0.9--2.0) & 0.6(0.3--1.1) & \nl
72 & 16:27:37.3 & $-$24:42:38.5 & 2.4 & 0.16 & 1.5(0.4--15.4) & 51.4(50.5--53.6) & 3.2(0.7--14.1) & 0.4(0.2--0.6) & 0.7(0.09--99.6) & \nl
73 & 16:27:37.6 & $-$24:36:04.4 & 1.1 & 0.27 & 5.3($>$0.5) & 50.8(50.3--53.2) & 3.3(0.8--14.0) & 0.4(0.3--0.6) & 0.3(0.1--71.4) & \nl
74  &  16:27:38.2  &  $-$24:30:42.6  & 0.6 &  1.71  &  1.6(1.0--2.7)  &  52.0(51.8--52.4)  &  1.4(1.1--2.1)  & 2.7(2.1--3.3) & 2.8(1.7--6.4) &  flare \nl
 &   &   &   &   &  1.3(0.9--2.0)  &  51.7(51.4--52.0)  &   & 1.0(0.8--1.2) & 1.3(0.7--2.9) &  quiescent \nl
75 & 16:27:38.3 & $-$24:36:58.4 & 0.2 & 12.3 & 1.7(1.5--1.9) & 52.7(52.6--52.8) & 2.0(1.8--2.2) & 12.5(11.8--13.2) & 14.7(12.1--17.5) & \nl
76 & 16:27:38.6 & $-$24:38:39.1 & 0.5 & 2.83 & 2.0(1.5--2.5) & 51.5(51.4--51.6) & 0.5(0.3--0.7) & 1.9(1.7--2.1) & 1.0(0.8--1.3) & \nl
77  &  16:27:39.0  &  $-$24:40:20.5  & 0.4 &  7.38  &  5.5(3.2--11.2)  &  53.4(53.3--53.7)  &  8.2(6.9--10.1)  & 144(130--158) & 133.0(93.5--223.6) &  flare \nl
 &   &   &   &   &  3.8(2.5--7.1)  &  52.4(52.2--52.6)  &   & 8.5(7.6--9.5) & 9.7(5.9--17.5) &  quiescent \nl
78 & 16:27:39.5 & $-$24:39:15.5 & 0.1 & 66.7 & 3.1(2.9--3.4) & 53.2(53.2--53.3) & 1.6(1.5--1.6) & 92.4(90.1--94.7) & 62.1(57.7--66.7) & \nl
79\tablenotemark{d}    &  16:27:39.9  &  $-$24:43:15.3  & 0.7 &  9.36  &  $>$10.6  & 53.2(53.1--53.2) & 4.8(4.3--5.5) & 143(122--163) & 83.9(71.6--96.8) &  flare1 \nl
 &   &   &   &   &  3.5(2.0--13.0) & 53.3(53.0--53.5) &  & 74.4(55.8--93.0) & 71.0(38.7--124.9) & flare2 \nl
 &   &   &   &   &  1.9(1.6--2.3) & 52.9(52.7--53.0) &  & 12.5(11.6--13.5) & 20.9(15.3--28.9) &  quiescent \nl
80 & 16:27:41.5 & $-$24:35:37.7 & 0.9 & 0.54 & 1.2(0.5--2.8) & 51.6(51.0--52.7) & 2.5(1.4--4.2) & 0.5(0.4--0.6) & 1.0(0.3--14.4) & \nl
81 & 16:27:42.8 & $-$24:38:51.1 & 0.8 & 1.45 & 0.9(0.6--1.2) & 52.1(51.7--52.6) & 2.3(1.6--3.2) & 1.0(0.8--1.2) & 3.8(1.5--12.0) & \nl
82\tablenotemark{d} & 16:27:43.7 & $-$24:31:26.6 & 0.6 & 2.49 & $>$5.5 & 51.9(51.8--52.0) & 5.2(4.4--6.5) & 19.3(16.9--21.8) & 8.9(7.6--10.4) & \nl
83\tablenotemark{d} & 16:27:49.1 & $-$24:39:06.9 & 1.7 & 0.29 & $>$0.4 & 50.9(50.3--51.3) & 4.4(0.5--17.7) & 2.1(1.3--2.9) & 1.0(0.3--2.4) & \nl
84 & 16:27:50.4 & $-$24:31:49.3 & 1.0 & 1.64 & 2.8(1.2--7.5) & 52.1(51.6--53.1) & 6.4(3.9--15.4) & 3.2(2.7--3.7) & 4.0(1.5--42.9) & \nl
85 & 16:27:52.0 & $-$24:40:50.0 & 0.3 & 46.3 & 2.3(2.2--2.4) & 53.2(53.2--53.2) & 0.8(0.7--0.8) & 81.3(78.9--83.7) & 49.5(46.9--52.9) & \nl
86\tablenotemark{a}  &  16:27:54.7  &  $-$24:37:14.9  & 1.9 &  0.50  &  1.0(fixed)  &  52.7(52.4--53.0)  &  14.2(9.8--21.8)  & 0.8(0.6--1.0) & 15.7(7.3--30.8) & \nl
 &   &   &   &   &  5.0(fixed)  &  51.4(51.1--51.6)  &  6.6(3.8--12.0)  & 1.2(0.9--1.5) & 1.1(0.6--1.8) & \nl
87 & 16:27:55.3 & $-$24:28:39.4 & 2.2 & 0.28 & 1.1(0.1--3.7) & 51.2(50.6--56.4) & 1.3(0.2--4.0) & 0.3(0.2--0.4) & 0.5($>$0.1) & \nl
\enddata \tablecomments{
(1):~Source  number.  
(2),(3):~Source coordinate in the equinox 2000. 
(4):~Position uncertainty [arc-sec] with the 90\% confidence limit. 
(5):~Photon counts ks$^{-1}$. 
(6),(7),(8),(9),(10):~Plasma temperature [keV], log(Emission Measure
[cm$^{-3}$]), absorption column density [10$^{22}$ cm$^{-2}$], observed
 flux [10$^{-14}$ ergs s$^{-1}$ cm$^{-2}$] in 0.5--10.0 keV,
 and absorption-corrected luminosity [10$^{29}$ ergs s$^{-1}$] in 0.5--10.0
 keV, respectively.  Parentheses indicate  the 90\% confidence limits.
(11):~Phases for the spectral fit. Flare 1 and 2 
indicate 2 flares, which are separately treated (see text).}
\tablenotetext{a}{Spectral parameters are obtained with fixed temperatures of 1 keV and 5 keV (see text).}  
\tablenotetext{b}{No spectral fit is done due to limited flux ($<$0.06 cts ks$^{-1}$).}
\tablenotetext{c}{Abundances are free parameters (see \S4.8.1 and Figure
 8).} 
\tablenotetext{d}{The best-fit value of $kT$ is not determined (larger
 than 20 keV), hence we
 assume 20 keV temperature for the estimation of the other parameters.}
\tablenotetext{e}{Foreground star. The distance is assumed to be 60 pc
 (Festin 1998).} 
\tablenotetext{f}{Pile-up effect is not corrected.}
\tablenotetext{g}{Errors are not constrained.}
\end{deluxetable}
}

{\footnotesize
\begin{deluxetable}{lcccclccll}
\tablewidth{0pt}
\tablenum{2}
\tablecaption{Identifications of the X-ray sources \label{tab:id}}
\tablehead{ \colhead{No.} & \colhead{BKLT} &
\colhead{PSPC} & \colhead{HRI} & \colhead{ASCA} & \colhead{Other names} &
\colhead{Class} & \colhead{$L_{\rm bol}$} &
\multicolumn{2}{c}{Reference} \nl \colhead{(1)} &
\colhead{(2)} & \colhead{(3)} & \colhead{(4)} & \colhead{(5)} &
\colhead{(6)} & \colhead{(7)} & \colhead{(8)} & \colhead{Class} & \colhead{$L_{\rm bol}$}}
\startdata
1 & 162644$-$243447 & \nodata & \nodata & \nodata & WL12/YLW2/GY111/VSSG30/BBRCG4 & I & 6.4 & 1,2,3,4,5 & 3 \nl
2 & 162648$-$242836 & \nodata & \nodata & \nodata & WL2/GY128/BBRCG6 & II & \nodata & 1 & \nl 
3 & \nodata & \nodata & \nodata & \nodata & \nodata & \nodata & \nodata & & \nl  
4 & \nodata & \nodata & \nodata & \nodata & \nodata & \nodata & \nodata & & \nl  
5 & \nodata & \nodata & \nodata & \nodata & \nodata & \nodata & \nodata & & \nl  
6 & \nodata & \nodata & \nodata & \nodata & \nodata & \nodata & \nodata & & \nl  
7 & 162654$-$242622 & 33 & \nodata & \nodata & VSSG5/GY153 & II & 1.2 & 1 & 5 \nl
8 & 162657$-$243538 & 34 & \nodata & \nodata & WL21/IRS23/GY164/BBRCG/17/CRBR50/SKS3-23 & I$_c$ & 0.0 & 4 & 6 \nl
9 & 162659$-$243556 & \nodata & \nodata & \nodata & WL14/GY172/BBRCG18/CRBR52/SKS3-25 & I$_c$ & \nodata & 4 & \nl 
10 & 162659$-$243458 & \nodata & \nodata & \nodata & WL22/IRS27/GY174/CRBR53 & I$_c$ & 15.7 & 3,4 & 3 \nl
11 & \nodata & \nodata & \nodata & \nodata & \nodata & \nodata & \nodata & & \nl  
12 & 162703$-$242615 & \nodata & \nodata & \nodata & GY188 & ? & \nodata & & \nl  
13 & 162704$-$242830 & \nodata & \nodata & \nodata & WL1/IRS29/GY192/VSSG31/BBRCG20/CRBR55 & I$_c$ & 1.1 & 3 & 3 \nl
14 & 162704$-$242716 & \nodata & \nodata & \nodata & GY195 & ? & \nodata & & \nl  
15 & \nodata & \nodata & \nodata & \nodata & \nodata & \nodata & \nodata & & \nl  
16 & 162705$-$242619 & \nodata & \nodata & \nodata & GY203/IRS30 & ? & \nodata & & \nl  
17 & 162706$-$244147 & \nodata & \nodata & \nodata & GY204/SKS1-26 & ? & \nodata & & \nl  
18 & 162706$-$243814 & \nodata & \nodata & C5 & WL17/GY205/BBRCG22/CRBR58/SKS3-27 & I$_c$ & \nodata & 5 & \nl 
19 & \nodata & \nodata & \nodata & \nodata & \nodata & \nodata & \nodata & & \nl  
20 & \nodata & \nodata & \nodata & \nodata & \nodata & \nodata & \nodata & & \nl  
21 & 162709$-$243408 & 37 & \nodata & 7 & WL10/GY211/BBRCG24/CRBR59/SKS1-27 & II & 1.4 & 2 & 5 \nl
22 & 162709$-$244022 & \nodata & \nodata & \nodata & GY213/SKS3-28 & ? & \nodata & & \nl  
23 & 162709$-$243718 & \nodata & \nodata & 8 & EL29/YLW7/GY214/BBRCG23/CRBR60/SKS1-28 & I & 36.3 & 1,2,3,4,5 & 3 \nl
24 & 162711$-$244046 & \nodata & \nodata & \nodata & GY224/SKS3-30 & I$_c$ & 0.3 & 2,4,5 & 3 \nl
25 & 162711$-$243832 & \nodata & \nodata & \nodata & WL19/YLW11/GY227/BBRCG28/CRBR65/SKS3-31 & I$_c$ & 0.1 & 5,6 & 6 \nl
26 & 162715$-$242640 & \nodata & \nodata & \nodata & IRS34/YLW10/GY239/SKS3-36 & I$_c$ & 0.8 & 3 & 3 \nl
27 & 162715$-$243054 & \nodata & F15 & \nodata & GY238/BBRCG32/CRBR70/SKS3-35/IRS35 & II & 0.4 & 7 & 7 \nl
28 & \nodata & \nodata & \nodata & \nodata & \nodata & \nodata & \nodata & & \nl  
29 & 162715$-$243843 & 39 & F16 & \nodata & WL20/GY240/BBRCG31/CRBR68/SKS1-31 & I$_c$ & 1.3 & 1,4 & 7 \nl
30 & \nodata & \nodata & \nodata & \nodata & \nodata & \nodata & \nodata & & \nl  
31 & \nodata & \nodata & \nodata & \nodata & \nodata & \nodata & \nodata & & \nl  
32 & \nodata & \nodata & \nodata & \nodata & \nodata & \nodata & \nodata & & \nl  
33 & 162717$-$242856 & \nodata & \nodata & \nodata & IRS37/YLW12A/GY244/VSSG26/BBRCG36/CRBR71/SKS1-32 & II? & \nodata & 2 & \nl 
34 & 162718$-$242853 & 40 & \nodata & \nodata & WL5/IRS38/GY246/VSSG26/BBRCG37/CRBR73/SKS1-33 & III & 0.2 & 2 & 3 \nl
35 & 162718$-$243915 & \nodata & \nodata & \nodata & GY245/BBRCG39/CRBR72/SKS3-37 & I$_c$ & 0.1 & 2,4 & 3 \nl
36 & 162718$-$242906 & \nodata & \nodata & C6 & WL4/IRS39/GY247/VSSG26/BBRCG40/CRBR75/SKS1-34 & II & 1.6 & 1 & 5 \nl
37 & \nodata & \nodata & \nodata & \nodata & \nodata & \nodata & \nodata & & \nl  
38 & 162719$-$242844 & \nodata & \nodata & \nodata & WL3/IRS41/GY249/VSSG26/CRBR77/SKS3-38 & II & 0.05 & 1 & 6 \nl
39 & \nodata & \nodata & \nodata & \nodata & \nodata & \nodata & \nodata & & \nl  
40 & 162719$-$244139 & 41 & F17 & 9A & SR12AB/ROXs21/GY250/CRBR76/SKS1-35/IRS40 & III & 1.1 & 1 & 7 \nl
41 & \nodata & \nodata & \nodata & \nodata & \nodata & \nodata & \nodata & & \nl  
42 & \nodata & \nodata & \nodata & \nodata & \nodata & \nodata & \nodata & & \nl  
43 & 162721$-$244142 & \nodata & \nodata & \nodata & IRS42/YLW13B/GY252/CRBR79/SKS1-36 & II & 0.8 & 1 & 6 \nl
44 & 162721$-$242953 & \nodata & \nodata & 10 & WL6/YLW14/GY254/CRBR80/SKS3-40 & I & 1.0 & 1,2,4,5,6 & 6 \nl
45 & 162722$-$242939 & \nodata & \nodata & \nodata & GY256/BBRCG43/CRBR81/SKS3-41 & I & \nodata & 2 & \nl 
46 & \nodata & \nodata & \nodata & \nodata & \nodata & \nodata & \nodata & & \nl  
47 & \nodata & \nodata & \nodata & \nodata & \nodata & \nodata & \nodata & & \nl  
48 & 162724$-$242929 & \nodata & \nodata & \nodata & GY257/CRBR83/SKS3-43 & ? & \nodata & & \nl  
49 & 162724$-$244147 & \nodata & \nodata & \nodata & GY258/CRBR84/SKS3-42 & ? & \nodata & & \nl  
50 & 162724$-$242935 & \nodata & \nodata & \nodata & GY259/CRBR86/SKS3-45 & ? & \nodata & & \nl  
51 & \nodata & \nodata & \nodata & \nodata & \nodata & \nodata & \nodata & & \nl  
52 & 162726$-$242555 & \nodata & \nodata & \nodata & GY264/SKS1-37 & F & \nodata & 8 & \nl 
53 & 162726$-$243923 & \nodata & \nodata & \nodata & GY262/BBRCG45/CRBR88/SKS3-47 & II & 1.0 & 2 & 5 \nl
54 & 162726$-$244051 & 43 & F21 & T\tablenotemark{a} & IRS43/YLW15A/GY265/CRBR89/SKS3-49 & I & 13.9 & 1,2,3,4,5 & 7 \nl
55 & 162727$-$243217 & \nodata & \nodata & \nodata & GY266/SKS3-5 & ? & \nodata & & \nl  
56 & 162727$-$243116 & 44 & F22 & \nodata & VSSG25/WL13/GY267/VSSG25/EL31/SKS1-38 & II & 0.6 & 1 & 7 \nl
57 & 162728$-$243934 & 45 & G\tablenotemark{b} & \nodata & IRS44/YLW16A/GY269/BBRCG49/SKS3-51 & I & 8.9 & 1,2,3,4,5 & 3 \nl
58 & \nodata & \nodata & \nodata & \nodata & \nodata & \nodata & \nodata & & \nl  
59 & 162728$-$242721 & \nodata & \nodata & \nodata & VSSG18/IRS45/GY273/VSSG18/EL32/SKS1-39 & II & 0.9 & 1 & 5 \nl
60 & 162729$-$243917 & \nodata & \nodata & 9B & IRS46/YLW16B/GY274/BBRCG52/SKS3-54 & I & 0.8 & 1,2,3,4,5 & 3 \nl
61 & 162730$-$243336 & \nodata & \nodata & \nodata & \nodata & ? & \nodata & & \nl  
62 & \nodata & \nodata & \nodata & \nodata & \nodata & \nodata & \nodata & & \nl  
63 & 162730$-$242744 & \nodata & \nodata & \nodata & VSSG17/IRS47/GY279/EL33/SKS1-40 & II & 1.8 & 1 & 5 \nl
64 & 162731$-$243402 & 47 & \nodata & \nodata & GY283/BBRCG56 & ? & \nodata & & \nl  
65 & \nodata & \nodata & \nodata & \nodata & \nodata & \nodata & \nodata & & \nl  
66 & 162732$-$243323 & \nodata & \nodata & \nodata & GY289/BBRCG60 & ? & \nodata & & \nl  
67 & 162732$-$243235 & 48? & \nodata & \nodata & GY291/BBRCG61 & ? & \nodata & & \nl  
68 & \nodata & \nodata & \nodata & \nodata & \nodata & \nodata & \nodata & & \nl  
69 & 162733$-$241115 & 49 & F27 & \nodata & GY292/CRBR90 & II & 2.2 & 2 & 7 \nl
70 & 162735$-$243834 & \nodata & \nodata & \nodata & GY295/SKS1-44 & III & 0.4 & 5 & 5 \nl
71 & \nodata & \nodata & \nodata & \nodata & \nodata & \nodata & \nodata & & \nl  
72 & 162737$-$244238 & \nodata & \nodata & \nodata & GY301/SKS3-59 & ? & \nodata & & \nl  
73 & \nodata & \nodata & \nodata & \nodata & \nodata & \nodata & \nodata & & \nl  
74 & 162738$-$243043 & \nodata & \nodata & \nodata & IRS50/GY306/SKS1-47 & III & 0.6 & 1 & 5 \nl
75 & 162738$-$243658 & 50 & F29 & \nodata & IRS49/GY308/SKS1-48 & III & 1.5 & 5 & 7 \nl
76 & 162738$-$243838 & \nodata & \nodata & \nodata & GY310/SKS1-49 & BD & 0.08 & 9 & 5 \nl
77 & 162738$-$244019 & \nodata & \nodata & \nodata & GY312/SKS3-61 & ? & \nodata & & \nl  
78 & 162739$-$243914 & 51 & F30 & \nodata & GY314/SKS1-50 & II & 1.1 & 2 & 7 \nl
79 & 162739$-$244316 & \nodata & \nodata & T\tablenotemark{a} & IRS51/YLW45/GY315/SKS3-63 & I$_c$ & 2.2 & 1,2,3,4 & 3 \nl
80 & 162741$-$243537 & \nodata & \nodata & \nodata & GY322 & ? & \nodata & & \nl  
81 & 162742$-$243849 & \nodata & \nodata & \nodata & GY326/SKS1-52 & BD$_c$ & 0.4 & 9 & 5 \nl
82 & \nodata & \nodata & \nodata & \nodata & \nodata & \nodata & \nodata & & \nl  
83 & \nodata & \nodata & \nodata & \nodata & \nodata & \nodata & \nodata & & \nl  
84 & \nodata & \nodata & \nodata & \nodata & \nodata & \nodata & \nodata & & \nl  
85 & 162752$-$244049 & 54 & G\tablenotemark{b} & T\tablenotemark{a} & ROXs31/IRS55/GY380 & III & 0.9 & 1 & 3 \nl
86 & \nodata & \nodata & \nodata & \nodata & \nodata & \nodata & \nodata & & \nl  
87 & 162755$-$242840 & \nodata & \nodata & \nodata & GY397 & ? & \nodata \nl    
\enddata 
\tablecomments{ (1):~Source number.  (2):~$JHK$ band counterpart
 (Barsony et al. 1997). (3):~$ROSAT$ PSPC sources (Casanova et
 al. 1995).  (4):~$ROSAT$ HRI sources (Grosso et al. 2000).  (5):~$ASCA$
 sources (Kamata et al. 1997).  (6):~Abbreviations: ~BBRCG (Barsony et
 al. 1989), CRBR (Comer\'on et al. 1993), EL (Elias 1978), GY (Greene \&
 Young 1992), IRS (Wilking, Lada, \& Young 1989), ROXs (Bouvier \&
 Appenzeller 1992), SKS (Strom, Kepner, \& Strom 1995), VSSG (Vrba et
 al. 1975), WL (Wilking \& Lada 1983), YLW (Young, Lada, \& Wilking
 1986).  (7):~Classifications: I:~Class I, I$_c$:~Class I candidate,
 II:~Class II, III:~Class III, BD:~brown dwarf, BD$_c$:~brown dwarf
 candidate, F:~foreground star.  Unclassified sources are indicated by
 question mark (?). (8):~Bolometric luminosity [$L_{\odot}$] assuming
 165 pc distance.}
 \tablenotetext{a}{Tsuboi et al. (2000)}
 \tablenotetext{b}{Grosso (2001)}
 \tablerefs{ (1) Andr\'e \& Montmerle 1994; (2)
 Casanova et al. 1995; (3) Chen et al. 1995; (4) Motte, Andr\'e, \& Neri
 1998; (5) Luhman \& Rieke 1999; (6) Chen et al. 1997; (7) Grosso et al. 
 2000; (8) Festin 1998; (9) Wilking, Greene, \& Meyer 1999.}
\end{deluxetable}
}

\begin{deluxetable}{ccccccc}
 \tablewidth{0pt} 
 \tablenum{3} 
 \tablecaption{Best-fit parameters of YLW 16A (No.57)  \label{tab:ylw16a}} 
 \tablehead{ \colhead{Phase\tablenotemark{a}} & \colhead{$kT$} &
 \colhead{log(EM)} & \colhead{$N_{\rm H}$\tablenotemark{b}} & \colhead{EW\tablenotemark{c}} &
 \colhead{Flux\tablenotemark{d}} & \colhead{$L$x\tablenotemark{e}} \nl
 & [keV] & [cm$^{-3}$]
 & [10$^{22}$cm$^{-2}$] & [eV] & [10$^{-12}$ergs s$^{-1}$ cm$^{-2}$] &
 [10$^{31}$ergs s$^{-1}$] }
 \startdata
1 & 3.6(2.7--5.8) & 52.8(52.7--53.0) & 4.7(4.5--4.8) & \nodata & 0.3(0.3--0.3) & 0.3(0.2--0.4) \nl
2 & 5.4(3.6--11.2) & 53.1(53.0--53.2) & 4.7 & \nodata & 0.8(0.7--0.9) & 0.6(0.5--0.8) \nl
3 & 6.1(4.6--8.5) & 53.4(53.4--53.5) & 4.7 & \nodata & 1.8(1.7--1.9) & 1.3(1.1--1.5) \nl
4 & 4.6(3.6--6.6) & 53.4(53.3--53.5) & 4.7 & \nodata & 1.3(1.2--1.4) & 1.1(0.9--1.3) \nl
5 & 3.7(2.8--5.4) & 53.1(53.0--53.2) & 4.7 & \nodata & 0.6(0.5--0.7) & 0.5(0.4--0.6) \nl
6 & 3.1(2.6--4.1) & 53.0(52.9--53.1) & 4.7 & \nodata & 0.3(0.3--0.3) & 0.3(0.2--0.4) \nl
7 & 10.7(7.8--18.9) & 54.1(54.0--54.1) & 4.7 & 146(44--248) & 10.9(10.3--11.5) & 6.8(6.3--7.4) \nl
8 & 11.8(9.2--17.6) & 54.5(54.4--54.5) & 4.7 & 95(23--167) & 26.9(25.8--28.0) & 16.7(15.7--17.8) \nl
9 & 8.1(6.7--10.4) & 54.4(54.4--54.4) & 4.7 & 92(25--160) & 20.1(19.3--20.9) & 13.3(12.3--14.5) \nl
 \enddata 
 \tablenotetext{a}{See Figure 9.}
 \tablenotetext{b}{$N_{\rm H}$ is assumed to be the same value in all the phases.}
 \tablenotetext{c}{Equivalent width for the 6.4 keV line.}
 \tablenotetext{d}{Observed flux in 0.5--10.0 keV.}
 \tablenotetext{e}{Absorption-corrected luminosity in 0.5--10.0 keV.}
\end{deluxetable}

{\small
\begin{deluxetable}{cccccccc}
 \tablewidth{0pt}
 \tablenum{4}
 \tablecaption{Best fit parameters of ROXs 21 (No.40) for the
 (thin-thermal + thin-thermal) model  \label{tab:roxs21}}
 \tablehead{ & & \colhead{$kT$} & \colhead{log(EM)} &
 \colhead{Abundance\tablenotemark{a}} & \colhead{$N_{\rm H}$\tablenotemark{a}} &
 \colhead{Flux\tablenotemark{b}} & \colhead{$L$x\tablenotemark{c}} \nl
 & & [keV] & [cm$^{-3}$] & [solar] &
 [10$^{22}$cm$^{-2}$] & [10$^{-14}$ergs s$^{-1}$ cm$^{-2}$] &
 [10$^{29}$ergs s$^{-1}$] }
 \startdata
 flare & soft & 0.84(0.76--1.04) & 52.5(52.3--52.6) &
 0.19(0.16--0.24) & 0.08(0.04--0.12) & 17.5(16.1--18.9) & 7.2(4.9--9.6) \nl
  & hard & 2.49(2.07--2.89) & 52.8(52.7--52.9) &
 0.19 & 0.08 & 51.9(48.9--54.9) & 19.1(17.2--22.3) \nl
 quiescent & soft & 0.83(0.78--0.88) & 52.5(52.4--52.6) &
 0.19 & 0.08 & 17.9(17.2--18.6) & 7.3(5.5--9.8) \nl
  & hard & 1.67(1.54--1.85) & 52.6(52.5--52.6) &
 0.19 & 0.08 & 24.3(23.4--25.2) & 9.2(7.9--9.9) \nl
 \enddata 
 \tablenotetext{a}{Abundance and $N_{\rm H}$ are assumed to be
 the same values in all the phases and components.}
 \tablenotetext{b}{Observed flux in 0.5--10.0 keV.}
 \tablenotetext{c}{Absorption-corrected luminosity in 0.5--10.0 keV.}
\end{deluxetable}
}